\documentclass[
  journal=pasa,
  manuscript=research-paper, 
  year=2025,
  volume=xx,
]{cup-journal}

\usepackage{microtype,siunitx,booktabs}
\usepackage{soul}
\usepackage[colorlinks=true,linkcolor=black, citecolor=black]{hyperref}
\usepackage{amsmath,amssymb, systeme}
\usepackage{xcolor}
\usepackage[normalem]{ulem} 
\usepackage{changes}

\usepackage{cancel} 

\defcitealias{YatesJones_2023}{Paper~I}
\newcommand{\PaperI}{\citetalias{YatesJones_2023}}

\newcommand{\stkout}[1]{\ifmmode\text{\cancel{\ensuremath{#1}}}\else\cancel{#1}\fi}

\title{CosmoDRAGoN II: Remnant Radio Galaxies in Group and Cluster Environments}

\author{Georgia S.C. Stewart} 
\affiliation{School of Natural Sciences, University of Tasmania, Private Bag 37, Hobart, TAS, 7001, Australia}
\alsoaffiliation{CSIRO Space and Astronomy, PO Box 1130, Bentley, WA 6102, Australia}
\email[G. S. C. Stewart]{georgia.stewart@utas.edu.au}

\author{Stanislav S. Shabala}
\affiliation{School of Natural Sciences, University of Tasmania, Private Bag 37, Hobart, TAS, 7001, Australia}

\author{Ross J. Turner}
\affiliation{School of Natural Sciences, University of Tasmania, Private Bag 37, Hobart, TAS, 7001, Australia}

\author{Patrick M. Yates-Jones}
\affiliation{School of Natural Sciences, University of Tasmania, Private Bag 37, Hobart, TAS, 7001, Australia}

\author{Martin G. H. Krause}
\affiliation{Centre for Astrophysics Research, University of Hertfordshire, College Lane, Hatfield, Herts AL10 9AB, UK}

\author{O. Ivy Wong}
\affiliation{CSIRO Space and Astronomy, PO Box 1130, Bentley, WA 6102, Australia}
\alsoaffiliation{International Centre for Radio Astronomy Research, University of Western Australia, 35 Stirling Highway, Crawley, Western Australia 6009, Australia}

\author{Chris Power}
\affiliation{International Centre for Radio Astronomy Research, University of Western Australia, 35 Stirling Highway, Crawley, Western Australia 6009, Australia}
\alsoaffiliation{ARC Centre of Excellence for All Sky Astrophysics (ASTRO 3D), Australia}

\author{Martin J. Hardcastle}
\affiliation{Centre for Astrophysics Research, University of Hertfordshire, College Lane, Hatfield, Herts AL10 9AB, UK}

\doi{xxxxx}

\received {dd Mmm YYYY}
\revised  {dd Mmm YYYY}
\accepted {dd Mmm YYYY}
\published{dd Mmm YYYY}


\begin{document}

\defcitealias{YatesJones_2023}{Paper I}

\begin{abstract}

Radio galaxy remnants are a rare subset of the radio-loud active galactic nuclei (RLAGN) population, representing the quiescent phase in the RLAGN lifecycle. Despite their observed scarcity, they offer valuable insights into the AGN duty cycle and feedback processes. Due to the mega-year timescales over which the RLAGN lifecycle takes place, it is impossible to observe the active to remnant transition in real-time. Numerical simulations offer a solution to follow the long-term evolution of RLAGN plasma. In this work, we present the largest suite (to date) of three-dimensional, hydrodynamic simulations studying the dynamic evolution of the active-to-remnant transition and explore the mechanisms driving cocoon evolution, comparing the results to the expectations of analytic modelling. Our results show key differences between active and remnant sources in both cluster environments and in lower-density group environments. We find that sources in low-density environments can remain overpressured well into the remnant phase. This significantly increases the time for the remnant lobe to transition to a buoyant regime. We compare our results with analytic expectations, showing that the long-term evolution of radio remnants can be well captured for remnants whose expansion is largely pressure-driven if the transition to a coasting phase is assumed to be gradual. We find that remnants of low-powered progenitors can continue to be momentum-driven for about 10 Myr after the jets switch-off. Finally, we consider how the properties of the progenitor influence the mixing of the remnant lobe and confirm the expectation that the remnants of high-powered sources have long-lasting shocks that can continue to heat the surrounding medium. 

\end{abstract}


\section{INTRODUCTION}

Pairs of radio-emitting jets launched via accretion processes in radio-loud active galactic nuclei (RLAGN) are among the most energetic objects in the Universe and are known to be an episodic phenomenon, operating on a duty cycle over the lifetime of the host galaxy. The most obvious evidence for episodic jet activity is shown by ``double-double'' radio galaxies (\citealp[DDRGS;][]{Schoenmakers_et_al_2000}) in which two pairs of radio lobes are observed; the inner pair signalling recent jet production and the outer pair leftover from a past cycle of activity. Further direct evidence is supplied by the ripples and multiple X-ray cavities seen in the Perseus \citep{Boehringer_1993,Fabian_2003} and Virgo clusters \citep{Forman_2005}. Indirect evidence of a strong correlation between quiescent star formation and the presence of active RLAGN in local, massive galaxies implies prolonged radio-loud phases where the emitted energy is sufficient to prevent the cooling of halo gas \citep{Best_2005, Best_2007, Bari_i__2017, Sabater_2019}. From a theoretical perspective, episodic behaviour is expected and necessary for a self-regulating feedback loop to achieve a balance between cooling, infalling gas, and the heating, and redistribution of that gas \citep[e.g.][]{Kawata_2005, Shabala_Alexander_feedback_2009, Novak_2011, Pope_2011, Raouf_2017}. \\

\begin{figure*}      
    \includegraphics[trim={2.5cm 0.5cm 2.5cm 1.3cm}, width=0.70\linewidth]{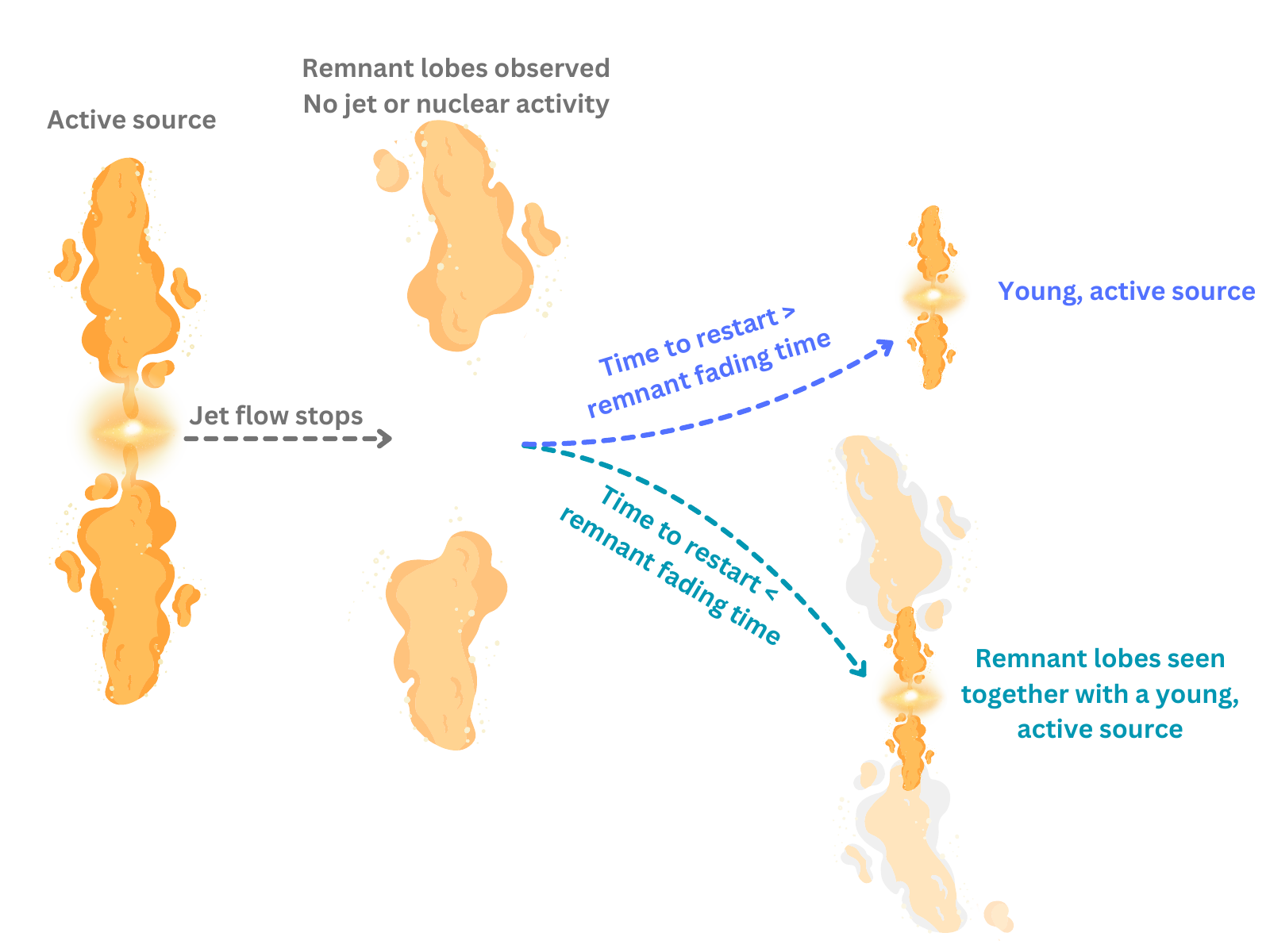}
    \caption{A schematic of the AGN lifecycle from an active source (left), to a remnant source (middle) to restarted sources (right). If the time for the remnant lobes to fade below the detectable limit is longer than the time for the nuclear activity to reignite, then remnant lobes may be seen together with a newly restarted, young radio source (bottom right). Else, a young radio source may be seen with no indication of past activity (top right).}
    \label{fig:AGNlifecycle}
\end{figure*}

The lifecycle of an AGN is often broken down into four stages in the literature; the youth stage, the evolved active phase, the remnant phase, and the restarted phase. A schematic of the AGN lifecycle is shown in Fig. \ref{fig:AGNlifecycle}. The youth stage is often identified with Compact Steep Spectrum (CSS) or Gigahertz Peaked Spectrum (GPS) sources \citep{O_Dea_1998}. As they pass through the host galaxy, RLAGN jets drive gas outflows \citep{Morganti_2005, Morganti_2013, Mahony_2015, Santoro_2020}, create turbulence, and shock-heat the ISM, affecting star formation rates (either by enhancing it as in \citealp{Croft_2006, Dugan_2017, Nesvadba_2020, Zovaro_2020}, or by suppressing it as in \citealp{Nesvadba_2010, Nesvadba_2011, Morganti_2013, Mandal_2021}).\\

If the RLAGN jets evolve to larger scales  \footnote{The vast majority of AGN jets do not make it to large sizes. Both modelling and observations suggest an abundance of shorter-lived, smaller-scale sources \citep{Fanti_1995, Giroletti_2009, An_Baan_2012, Hardcastle_2019, Shabala_2020}}, the jet-inflated lobes can modify the thermal state of the surrounding environment, quenching otherwise rapid cooling at the centres of galaxy clusters \citep{Churazov_2001, Shabala_sound_waves_2009, Mittal_2009, Fabian_2012, Yang_Reynolds_2016}. The effects of `jet-mode’ feedback are now explicitly included in all modern cosmological galaxy formation models \citep{Sijacki_2007, Schaye_2014, Weinberger_2016, Dav__2019, Li_2021, Bird_2022, Villaescusa_Navarro_2023}. Cosmological models implement jet feedback in a variety of ways ranging from isotropic energy-momentum injection to collimated beams. However, due to the hugely dynamic range that would need to be covered, all cosmological models struggle to capture important features of real jets including relativistic jet speeds and initially wide opening angles. On the other hand, dedicated jet simulations can miss the complexity of cosmological environments. A few works have brought the two together including \cite{Mendygral_2012} and the CosmoDRAGoN simulation suite of \cite{YatesJones_2023}.\\

Following an active episode, the remnant stage (the focus of this work) will begin where the nuclear activity ceases or substantially reduces \citep{Sabater_2019}, the jet outflow stops, and the remnant lobes are left to  fade and rise buoyantly out of the central gravitational well \citep{Churazov_2001, Reynolds_2005}. As they rise, they entrain and uplift large quantities of low entropy gas to large radii \citep{Chen_Heinz_Ensslin_2019}. Finally, a restarted phase may occur where the nuclear activity reignites and a new jet pair may be observed \citep{Brocksopp_2011, Orru_2015, Nandi_2019, Mahatma_2023}. \\

Compared to the active phase of RLAGN evolution, the remnant phase is considerably less-well studied. Only a handful of studies have presented small observed samples of remnant candidates identified based on the assumptions about the morphological and spectral properties of aged radio plasma \citep[e.g.][]{Murgia_2011, Mahatma_2018, Jurlin_2020, Jurlin_2021, Quici_2021, Dutta_2023}. The numerical modelling that has concentrated on the large-scale evolution of the remnant phase has been mostly focused on the energy transfer between the remnant lobe and surrounding medium for powerful (Q$_{\rm{j}} = 10^{38-49}$ W), large-scale (>100 kpc), and, often, FR-II-like sources in smooth, radially symmetric environments \citep{Zanni_2005,Perucho_2011, Perucho_2014, Walg_2014, English_2019, Chen_Heinz_Ensslin_2019}. Similarly, analytical models have focused on describing the transition of powerful, strongly overpressured active sources \citep[e.g.][]{Kaiser_Cotter_2002, Hardcastle_2018}. These studies describe a system where, after the jets switch off, the radio lobes continue to rise through the ambient atmosphere, potentially driven by forces beyond simple buoyancy \citep[e.g.][]{Perucho_2014, English_2019}. During the transition from active to remnant source, a significant portion of the lobe energy may remain contained within the lobes at the end of the active phase \citep{English_2019}. Various mechanisms for energy transfer have been proposed. Entropy studies of the surrounding environment by \citet{Zanni_2005} highlight bow shock heating as the primary driver. In contrast, \citet{Chen_Heinz_Ensslin_2019} argue that the thermal exchange between uplifted, low-entropy gas and the ambient medium plays a significant role in heating cool-core systems. The overall paucity of remnant studies is unfortunate, as these systems hold key information about the late-stage evolution of RLAGN, the efficiency of energy injection into the surrounding medium, and the timescales over which feedback persists \citep{Turner_2018, Quici_2025}. \\

In a previous work, \citep[][hereafter \PaperI]{YatesJones_2023} we presented the first results from the CosmoDRAGoN project; a large suite of AGN jet simulations carried out in environments derived from cosmological simulations. In that paper, the dynamic properties and spatially-resolved spectral properties of active sources were primarily considered for simulations spanning both low and high jet kinetic powers corresponding to FR-I and FR-II morphologies. The evolution of remnants was only briefly considered.\\

The present paper extends the results of \PaperI ~by looking more closely at the transition from the active to remnant phase. We choose to use the CosmoDRAGoN suite of simulations, as natural inhomogeneities in a dynamic environment are likely to become important for the spatial distribution of remnant plasma and the timescales over which instabilities develop. Using these more complex environments, we explore the mechanisms driving lobe evolution after the active phase ends and compare the results from our numerical simulations to analytical expectations. In this paper, our results are based on data obtained from the raw hydrodynamic outputs. A study of the spectral properties of the simulations is deferred to the companion paper \cite{Stewart_2025}. The current paper is organised as follows: Section \ref{sec:methods} describes the computational setup and input parameters for our simulation suite. In Section \ref{sec:results} we present our findings, and in Section \ref{sec:discussion}, we provide a discussion of the results including comparisons to analytic modelling. Finally, in Section \ref{sec:conclusions} we summarise the work and present our conclusions.

\section{SIMULATION SETUP AND PARAMETERS}
\label{sec:methods}

To compile our simulation suite, we take five existing active simulations (i.e. remnant progenitors) from the CosmoDRAGoN simulation catalogue presented in \PaperI ~and continue their evolution in the remnant phase. In Section \ref{subsec:numerical_methods} we outline the numerical techniques used for all simulations. We then describe the method for environment selection and jet injection in Sections \ref{subsec:environment} and \ref{subsec:jet_injection} respectively. Because this work models the transition from active to remnant radio sources, we explain the process to terminate the jet fluid and evolve the remnant in Section \ref{subsec:jet_termination}. Finally, we describe our choice of remnant progenitor properties in Section \ref{subsec:sim_run_codes}.

\subsection{Numerical methods}
\label{subsec:numerical_methods}

All active and remnant simulations used in this work are run using a modified version of the publicly-available \textsc{pluto} code version 4.3 \footnote{http://plutocode.ph.unito.it/}. The numerical techniques employed follow those of \cite{Yates_Jones_2021} and \PaperI. The fluid variables are related by the Taub-Mathews equation of state \citep{Taub_1948, Mathews_1971, Mignone_McKinney2007} and we solve the Eulerian equations of relativistic hydrodynamics using the \texttt{hllc} Riemann solver. We use a linear reconstruction scheme and a Courant-Friedrichs-Lewy (CFL) number of 0.3. A second-order Runge-Kutta integration progresses the simulation in time and a \textsc{minmod} flux limiter is used to prevent spurious oscillations in the presence of shocks and discontinuities. We adopt the standard $\Lambda$CDM cosmology, which is consistent with the simulation catalogue that our environments are derived from (see Section \ref{subsec:environment} below). The parameters are taken from the \textit{Planck} mission and are $\Omega_{\rm{M}} = 0.307$, $\Omega_{\rm{B}} = 0.048$, $\Omega_{{\Lambda}} = 0.693$ \citep{Plank_Collab_2016}. \\

All simulations are run on a static, three-dimensional Cartesian domain (X, Y, Z) with maximum physical dimensions: $400$ kpc $\times 400$ kpc $\times 400$ kpc (-200 kpc to +200 kpc in all dimensions). The grid extents for a given simulation are chosen based on the expected size of the simulation (some simulations in this work do not require the full 400 kpc grid) but the resolution remains consistent across all runs.\\

All three dimensions are separated into five resolution patches such that the finest resolution is only used where it is needed (i.e. near the jet injection region). The inner ($-2.5$ kpc, $2.5$ kpc) patch is uniformly covered by 100 cells giving a resolution of 0.05 kpc/cell. In the intermediate  ($\pm 2.5$ kpc, $\pm 10$ kpc) region, a geometrically stretched grid is applied over 100 cells with a stretching ratio of $1.0076$. This gives a resolution of  0.08 kpc/cell at 10kpc. In the outer ($\pm 10$ kpc, $\pm 200$ kpc) patch, a stretching ratio of $1.0084$ is used with a resolution of 0.84 kpc/cell at 100 kpc. These stretching ratios result in a total grid cell count of 960 cells in each dimension. Following \PaperI, we impose outflow boundary conditions at all grid edges to avoid amplifying any initially small inhomogeneities present in our cosmological environments (see Section \ref{subsec:environment}).\\

The simulations of the active phase were run on the \textit{Gadi} facility provided by the National Computational Infrastructure, Australia. The follow-up remnant simulations were carried out using the \textit{kunanyi} facility provided by the Tasmanian Partnership for Advanced Computing. For our most computationally intensive model, the total computational time was approximately 790 000 CPU hours. The primary data products output from the simulation are the hydrodynamic quantities of density, pressure, velocity (in three dimensions), and a passive jet tracer value for each cell at each snapshot.

\subsection{Environment Implementation}
\label{subsec:environment}

The environment in which a radio galaxy evolves can have a significant impact on the morphology, dynamics, and observable properties such as the size, shape, brightness and spectral features of the radio emission \citep{Kaiser_Alexander_1997, Rodman_2018, Yates_Jones_2021}. The properties of this simulated environment are a particularly important consideration of the current study given that, in the absence of a driving momentum flux, remnant plasma will be more susceptible to the properties and inhomogeneities of the surrounding medium than an active source.\\
 
Some remnants have been associated with both cluster and group environments, while other remnants are considered to be isolated \citep{Murgia_2011, Jurlin_2021}. In our study, we choose to analyse simulations run in complex cluster and group-like environments. These complex environments are extracted from the pressure, temperature, and density profiles from the \textsc{the three hundred} project catalogue of re-simulated galaxy clusters with full-physics hydrodynamics \citep{Cui_2018}. The process for doing this is outlined in-depth in \PaperI, but we provide a brief overview here. \\

First, suitable halo candidates are selected from the \textsc{the three hundred} project catalogue. These are selected based on size, and overall stability, ensuring that there is no evidence of merger events at the epoch of interest. We use data from the $z=0$ snapshots are used throughout this paper.\\

Second, the pressure, density, momentum density, and force density are interpolated onto a three-dimensional Cartesian grid with a $1\rm\, kpc$/cell resolution such that total mass is conserved. The standard smooth particle hydrodynamics (SPH) `scatter' formalism, implemented in \textsc{sphtool}\footnote{https://bitbucket.org/at\_juhasz/sphtool/src/master/}, is used to smooth the environment quantities, such that the smoothed quantity $A_s$ is given as a function of position \textbf{$r$} and the SPH properties of $i$ particles,

\begin{equation}
    A_s(\mathbf{r}) \approx \sum_i m_i \frac{A_i}{\rho_i}W(|\mathbf{r}-\mathbf{r_i}|, h_i)
\end{equation}

with particle mass, density, and smoothing length $m_i$, $\rho_i$, and $h_i$ respectively. We use the $M_4$ kernel \citep{Monaghan_1985} for the smoothing function, $W(\mathbf{r}, h)$, as detailed further in \PaperI. The velocity and acceleration fields are respectively derived from the momentum and force fields. As \textsc{pluto} version 4.3 does not support self-gravity, \textsc{gadget-2} is used to calculate the gravitational acceleration necessary to generate gravitationally stable initial conditions. The gravitational acceleration is then interpolated using the same method as for the hydrodynamic quantities and initialised on the grid as a static field. This field cannot be evolved throughout the \textsc{pluto} simulations, resulting in a finite time before the gas density and gravitational potential fields dissociate from each other. However, stability tests conducted in \PaperI ~show the environments used in this work to be stable for at least $300$ Myr, which is longer than the timescales for which we simulate our remnant radio galaxies.\\

\begin{figure*}
    \centering\includegraphics[width=0.96\linewidth]{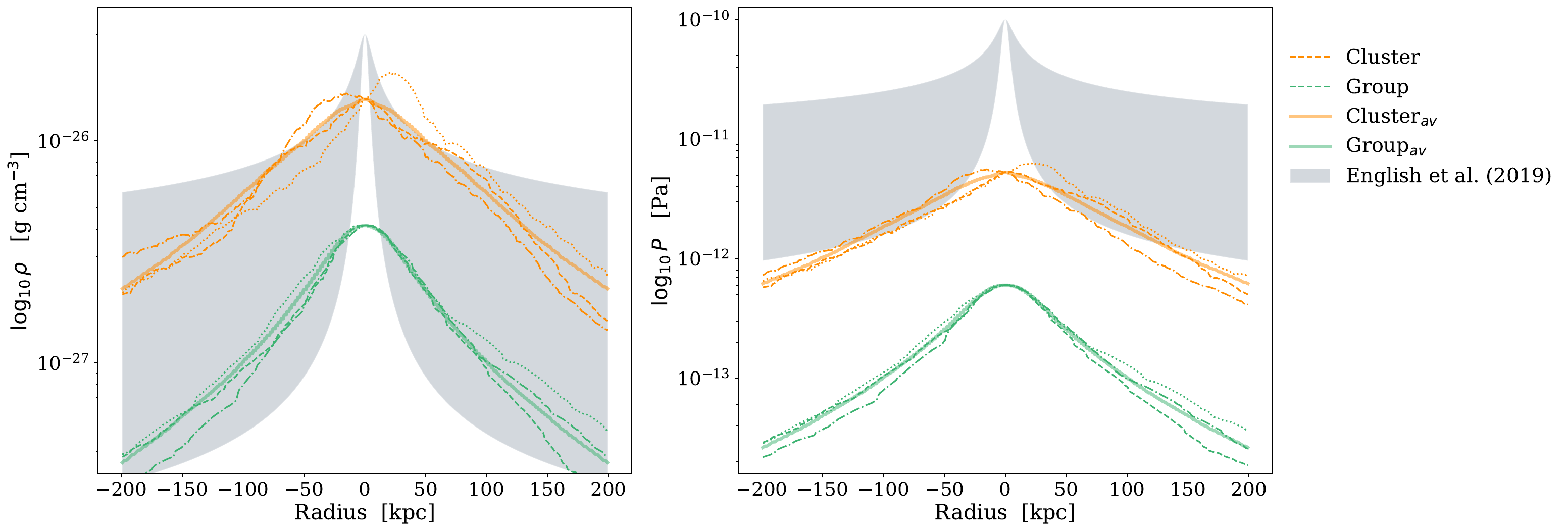}
    \caption{Environment density (left) and pressure (right) profiles are shown for the cosmological cluster (orange) and group (green) environments. Respectively, the dashed, dot-dashed and dotted lines show the radial evolution along the z-, x- and y- axes. The solid lines indicate the symmetric, radially averaged fit to the cosmological environments. For comparison, the range of environment profiles considered by the related work of \cite{English_2019} is shown by the shaded region.}
    \label{fig:environment comparison}
\end{figure*}

In the present work, we use remnant progenitor simulations run in halos 0003 and 0031 from cluster 002 in \textsc{the three hundred} project catalogue as the respective cluster- and group-like environments. These environments are chosen based on their long-term stability. A list of parameters for these halos is provided in Table \ref{tab:env_params} and in Fig. \ref{fig:environment comparison} we show the ambient density and pressure profiles.

\begin{table}
\begin{threeparttable}
\caption{Cosmological environment values for central density, central pressure, halo mass, virial radius, and average core radius.}
\label{tab:env_params}
\begin{tabular}{c l l} \toprule
\hline  & \textit{Cluster (halo 002-0003)} & \textit{Group (halo 002-0031)} \\ 
\hline
$\rho_{\rm{o}}$ (g/cm$^{-3}$) & $1.5 \times 10^{-26}$  &  $4.2 \times 10^{-27}$\\
$P_{\rm{o}}$ (Pa)             & $5.3 \times 10^{-12}$  & $6.0 \times 10^{-13}$ \\
$M_{\rm{halo}}$ (M$_{\odot}$) & $2.0 \times 10^{14} $ & $1.9 \times 10^{13}$ \\ 
$r_{\rm{halo}}$ (Mpc)         & $1.24$                 & $0.56$ \\
$r_{\rm{core}}$ (kpc)         & $59$                 & $42$ \\
\hline
\end{tabular}
\end{threeparttable}
\end{table}

\subsection{Jet Injection}
\label{subsec:jet_injection}

The remnant progenitors chosen from the CosmoDRAGoN simulation suite are injected as a pair of bipolar, conical outflows with half-opening angle $\theta_{\rm{j}}$. They are injected through a spherical, internal boundary condition of radius $r_{\rm{o}}=0.75$ kpc centred at the origin. Within $r < r_{\rm{o}}$, the pressure and density variables are continuously overwritten with the input jet density, $\rho_{\rm{j}}$, and pressure $p_{\rm{j}}$ which are calculated following

\begin{equation}
\rho (r < r_0) = 2\rho_{\rm{j}}\left[1+\left(\frac{r}{r_0}\right)^2\right]^{-1}
\end{equation}

\begin{equation}
P (r < r_0) = 2^\Gamma P_{\rm{j}}  \left( \frac{\rho(r)}{2\rho(r_{0})}\right)^\Gamma,
\end{equation}
where $r$ is the spherical radius from the grid origin and $\Gamma$ is the ideal gas adiabatic index calculated according to the Taub-Mathews equation of state. Within the jet injection cone, $\theta \leq \theta_j$, the fluid velocity is set to the jet velocity, $v_{\rm{j}}$.  For a relativistic regime, the one-sided jet kinetic power is related to the initial jet pressure, density, velocity, and cross-sectional area of the jet head at injection, $A_{\rm{j}}$ by

\begin{equation}\label{Eq. jet power}
Q_j = \left[\frac{\Gamma}{\Gamma - 1} P_{\rm{j}} \gamma^2 + \gamma(\gamma-1)\rho_{\rm{j}} c^2 \right]v_{\rm{j}} A_{\rm{j}},
\end{equation}
where $c$ is the speed of light in a vacuum and $\gamma = 1/\sqrt{1-v_{\rm{j}}^2/c^2}$ is the bulk flow Lorentz factor. 
For non-relativistic, strongly supersonic flows, equation \ref{Eq. jet power} reduces to the flux of kinetic energy density along the jet, 
\begin{equation}
Q = \frac{1}{2}\rho_{\rm{j}} v_{\rm{j}}^3A_{\rm{j} }.
\end{equation}

A passive tracer fluid is injected with the jet material, taking an initial value of 1 within the injection region and 0 elsewhere. 

\subsection{Jet Termination}
\label{subsec:jet_termination}

In this study, we focus on the evolution of simulated active radio jets into the remnant phase. To do this, we terminate the constant flow of injected material through the injection region and continue the simulation, following the evolution of the remnant lobes. When the jet flow is stopped, the fluid variables in the spherical injection region are no longer overwritten and assume similar values to their surrounding cells. To mitigate any adverse numerical affects after the jet is switched off, we include a `ramp-down' of the jet velocity over a time period of 0.05 Myr.\\

The timescales over which active sources transition to remnants are unclear. Optical and radio investigations suggest that the active phase can occur over long periods of $10-100$ Myr \citep{Parma_2007, Murgia_2011, Hardcastle_2019, Shabala_2020} as well as shorter periods of  $0.01-0.1$ Myr \citep{Murgia_2003, Polatidis_2003}. Once the jet flow is terminated, the remnant phase has been estimated to last anywhere from less than half of the active phase \citep{Parma_2007, Orru_2010, Murgia_2011}, to timescales comparable to \citep{Shulevski_2015}, or even longer than the active phase \citep{Brienza_2016}\footnote{These estimates are also dependent on how the remnant radio galaxy has been classified from observational data. The metrics for remnant classification are not clean-cut and we explore this further in the companion paper \cite{Stewart_2025}.} but recent modelling suggests that the fading time of remnant sources is rapid \citep{Yates_2018, English_2019, Shabala_2020}.  To explore a parameter space of varying active lifetimes and source sizes, each progenitor model (described below in Section \ref{subsec:sim_run_codes}) is terminated at three points: when the total source length has reached 20, 60, and 180 kpc. The total length of each lobe is approximately half the total source size, as all simulations show negligible asymmetries about the $x$-axis in the active phase. Terminating each model at 20, 60, and 180 kpc results in active lifetimes, $t_{\rm{on}}$, that span $0.5 - 100$ Myr. \\

\subsection{Simulation Parameters}
\label{subsec:sim_run_codes}

\begin{table*}[h]
\begin{threeparttable}
\caption{Parameters for the simulation runs. $Q_j$ is the one-sided kinetic jet power, $v_j$ is the initial jet velocity, $\theta_j$ is the half-opening angle, $t_{\rm{on}}$ is the time at which the remnant phase commences, and $t_{\rm{on}} + t_{\rm{off}}$ is the total simulation time. The run code of each model is given in the last column.}
\label{tab:sim_props}
\begin{tabular}{ c c c c c c c c } \toprule
\hline Environment & $Q_j$ & $v_j$ & $\theta_j$  & $t_{\rm{on}}$ & $t_{\rm{on}} + t_{\rm{off}}$& Run Code\\
 & (W) &  (c) & ($^o$) & (Myr) & (Myr)& \\ 
\hline         

\text {Cluster} &    $10^{36}$ & 0.01 & 25   & 5 &  100 & Q36-v01-a25-C20\\
                &             &      &      & 22 & 120 &Q36-v01-a25-C60\\
                &              &      &      & 100 & 200 &Q36-v01-a25-C180\\
\hline
                &    $10^{38}$ & 0.98 & 25   & 1.4 &  100& Q38-v98-a25-C20\\
                &              &      &      & 7.5 & 110 &Q38-v98-a25-C60\\
                &              &      &      & 56  & 155 &Q38-v98-a25-C180\\
\hline
                &    $10^{38}$ & 0.98 & 7.5  & 0.5 & 100 &Q38-v98-a7.5-C20\\
                &              &      &      & 1.5 & 95 &Q38-v98-a7.5-C60\\
                &              &      &      & 15  & 110 &Q38-v98-a7.5-C180\\
\toprule
\text {Group}   &    $10^{36}$ & 0.01 & 25   & 6 & 110 &Q36-v01-a25-G20\\
                &            &        &      & 16 & 115 &Q36-v01-a25-G60\\
                &            &      &        & 100 & 200 &Q36-v01-a25-G180\\
\hline
                &    $10^{38}$ & 0.98 & 25   & 0.8 & 75 &Q38-v98-a25-G20\\
                &            &      &       &  3.8 & 70 &Q38-v98-a25-G60\\
                &            &      &       &  22.8 & 95& Q38-v98-a25-G180\\
 \hline
\end{tabular}
\end{threeparttable}
\end{table*}

Our goal is to include a range of large-scale progenitor morphologies for evolving our remnants. Our parameter space, including jet-injection properties and environment, is consistent with this aim. FR-I morphologies are thought to be the result of a combination of non-relativistic or mildly relativistic velocities (0.01 - 0.5c)\footnote{FR-I sources are thought to propagate at relativistic speeds on parsec scales initially but decelerate to mildly relativistic speeds on scales up to a few tens of kiloparsecs \citep{Giovannini_2001,Laing_2014}. Given our jet injection radius of 0.75 kpc, we assume that the transition from a relativistic to a non-relativistic regime has already occurred.},  low kinetic jet powers ($10^{36} - 10^{37}$ W), and wide half-opening angles (> 24$^{\rm{o}}$) while FR-II-like morphologies are often characterised by relativistic injection velocities, narrow half-opening angles (< 24$^{\rm{o}}$) and high kinetic powers (> $10^{37}$ W) 
\citep{Krause_2012, Massaglia_2016}.\\

We explore a parameter space of low ($Q_{\rm{j}} = 10^{36}$ W) and high ($Q_{\rm{j}} = 10^{38}$ W) jet kinetic powers, relativistic (0.98c) and sub-relativistic (0.01c) jet injection velocities, narrow ($\theta_{\rm{j}} = 7.5^{\rm{o}}$) and wide ($\theta_{\rm{j}} = 25^{\rm{o}}$) half-opening angles across a range of active times in both group and cluster environments. The parameters for all simulation runs are shown in Table \ref{tab:sim_props}. Each simulation is assigned an identification code of the form \textbf{Qaa-vbb-acc-Ddd} where \textbf{Qaa} denotes the jet power (in log W), \textbf{vbb} is the injected velocity (as a fraction of c; i.e. v98 is 0.98c), \textbf{acc} is the half-opening angle (in degrees), \textbf{D} denotes the type of environment (C for cluster or G for group), and \textbf{dd} is the total source length at the start of the remnant phase (in kpc). We consider a total of 15 simulations, nine of which are run in the cluster environment, and six in the group.


\section{RESULTS}
\label{sec:results}

We have performed a series of hydrodynamic simulations in realistic environments to study how active and remnant dynamics are affected by different environment and jet initial conditions. We are particularly interested in the evolving morphology of the remnant lobes, their deceleration, and transition to pressure balance with the ambient medium. In this section, we examine the key morphological features of all simulations in \ref{subsec:radio_source_morphologies}. We quantify the dynamic evolution of the radio lobes in Section \ref{subsec:lobe_dynamics} before tracking the evolution of the shocked material surrounding the lobes in Section \ref{subsec:shocks}.

\subsection{General Features of Radio Source Evolution}
\label{subsec:radio_source_morphologies}

\subsubsection{Active Phase}
\label{subsubsec:active_source_properties}

\begin{figure*}
    \centering
    \includegraphics[width=0.8\linewidth]{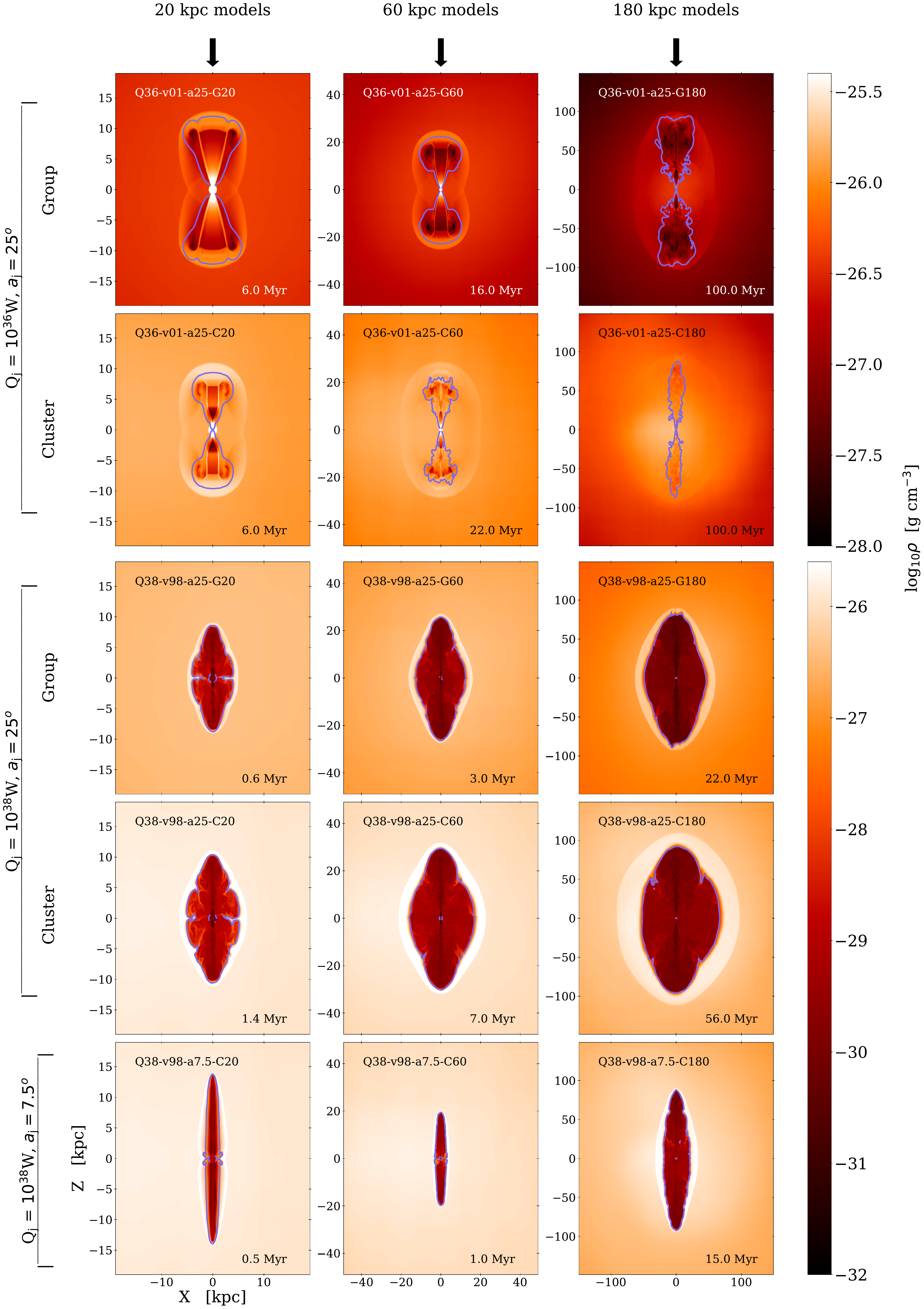}
    \caption{Mid-plane slices in the X-Z plane of the logarithmic density distributions for all models at the last active output. The total source age is displayed on each panel. From left to right, the columns show 20 kpc, 60 kpc, and 180 kpc switch-off simulations. The spatial scales have been adjusted across the three columns such that each simulation can be seen clearly. We have grouped the simulations such that low-power sources are shown in the top two rows and high-power sources in the bottom three rows.  Group simulations are shown in the panel immediately above the equivalent cluster simulation. The blue contours outline where the passive tracer is above the threshold of $10^{-4}$.}. 
    \label{fig:actives}
\end{figure*}

The morphology of each simulation is best understood by looking at the two-dimensional logarithmic density distributions. Mid-plane slices of this distribution for all simulations during the active phase are shown in Fig. \ref{fig:actives}. Each panel in Fig. \ref{fig:actives} corresponds to the last data output before the end of the active phase, $t_{\rm{on}}$. The values for $t_{\rm{on}}$ are given in table \ref{tab:sim_props}. The lobes are identified by the regions of underdense material (with respect to the local ambient medium) surrounded by a dense shell of hot, shocked ambient material. These lobes map well to the passive tracer quantity (the blue contour) which outlines all material with a tracer value greater than 10$^{-4}$. The amount of tracer material in the lobe is largely insensitive to this value, provided it is not greater than approximately 10$^{-2}$. Similar tracer cutoffs are used in other related work (\citealp{Hardcastle_Krause_2013, Yates_2018} and \PaperI). We consistently use a tracer cutoff of 10$^{-4}$ throughout this paper.\\

The slow, low-power simulations are shown in the top two rows of Fig. \ref{fig:actives}. All low-power simulations have similar jet kinetic power ($10^{36}$ W) and injection velocity ($0.01\rm{c}$) but differ in the environment they propagate into. The lobes of the low-power group (top row) and cluster (second row) simulations are filled with mildly ($\lesssim$ 0.5 dex) underdense, forward-flowing material, surrounded by an oval-shaped region of weakly shocked ambient gas. The magnitude of pressure in the shocked region is strongest at the lobe head and diminishes towards the equatorial regions. In the 180 kpc switch-off, cluster simulation, the weakly shocked material is nearly indistinguishable from the ambient medium (see simulation Q36-v01-a25-C180 in the far right panel of row two in Fig. \ref{fig:actives}).\\

The key difference between the low-power group and cluster simulations is the spatial distribution of the lobe plasma; group simulations inflate wider lobes. All six low-power simulations are injected as initially overdense, conical outflows and are left to collimate via a recollimation shock. Since the central density and pressure in the group environment is an order of magnitude less than in the cluster environment, simulations run in the group are collimated about $3$ times further downstream, and at a jet radius of around $3.5$ times larger than the equivalent cluster simulation. This is consistent with analytic theory, where the collimation length scale goes as $\rho_{\rm{x}}^{-1/2}$ \citep{Alexander_2006}. At later times, the higher pressures and densities of the cluster environment act to confine the inflating lobe material.\\

Our high-power simulations are shown in the bottom three rows of Fig. \ref{fig:actives}. All nine high-power simulations are run with an initial jet power of $Q_{\rm{j}} = 10^{38}$ W and injection velocity of $v_{\rm{j}} = 0.98$c characteristic of powerful radio galaxies. One set of simulations have a narrow half-opening angle of $7.5^o$ (bottom row), the other two have a wider half-opening angle of $25^o$, designed to produce FR-I sources \cite[e.g.][]{Krause_2012}.\\

All high-power sources are collimated within 5 kpc of injection and are characterised on large scales by significantly underdense (3-4 dex lower than the ambient medium) lobes. These are primarily filled by backflow from the strong terminal shock at the head of the lobe. Surrounding the lobes, the region of shocked ambient material (the `shocked shell') is roughly uniform in density. Such features are typical of overpressured outflows as seen in e.g. \cite{Komissarov_1998, Hardcastle_Krause_2013}.\\

For the wide jet opening angle simulations, the spatial distribution of the lobe plasma is similar regardless of environment (compare row three and four of Fig. \ref{fig:actives}). However, wide opening angle simulations in the cluster environment (row four in Fig. \ref{fig:actives}) consistently take twice as long as the group simulations to evolve to a certain size. For the narrow, high-power simulations, the lobes are thinner and the source takes only a third of the time to traverse the same distance as the wider angle simulation in the same environment. 

\begin{figure*}
    \centering
    \includegraphics[width=0.8\linewidth]{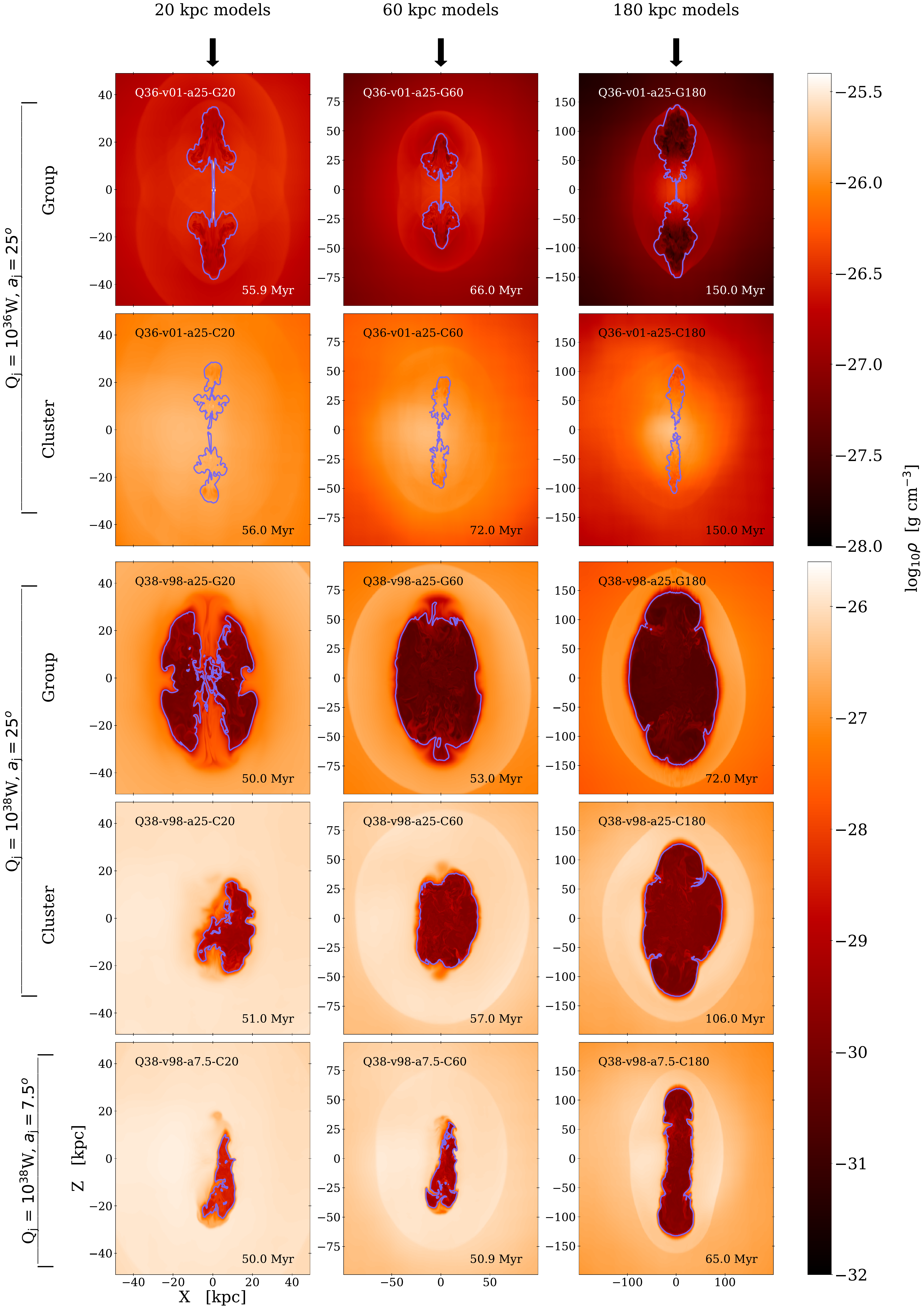}
    \caption{Mid-plane slices for all models 50 Myr after the remnant phase started. The layout of this figure is analogous to Fig. \ref{fig:actives}.}
    \label{fig:remnants}
\end{figure*}

\subsubsection{Remnant Phase}
\label{subsubsec:remnant_source_properties}

We allow each simulation to evolve for 100 Myr in the remnant phase, or until the shock front reaches the edge of the computational domain. Beyond this point, mass is no longer conserved on the grid, and the full interaction of the shock with the environment cannot be captured. Snapshots of our remnant sources 50 Myr post switch-off are given in Fig. \ref{fig:remnants}.
\\

In all sources, the shock front progressively separates from the lobes, becomes increasingly spherical, and weakens against the ambient medium. For some simulations (e.g. low-powered remnants in cluster environments), the shocked region has diminished such that it cannot be detected against the ambient medium 50 Myr after switch-off. For low-powered sources, the lobe separation increases and the inertia-driven material in the jet at switch-off continues along the evacuated channel, rising faster than the bulk lobe plasma. For our 20 kpc switch-off, low-power, cluster simulation (Q36-v01-a25-C20), the trailing jet material completely drills through the existing lobe and separates at $\sim \pm 20$ kpc from the grid origin, forming four main components (see the left-hand panel in row two of Fig. \ref{fig:remnants}). For larger low-power sources, the trailing jet material runs out of thrust before it traverses the lobe, and the lobes do not become fragmented at any point during the simulated time. \\

The evolution of high-power, fast simulations is different. The lobes are dominated by backflow, and this continues to mix in a single mass close to the core before the lobes become pinched due to large Rayleigh-Taylor instabilities. The trailing jet material quickly ($< 1$ Myr; less than the time between simulation outputs) reaches the lobe head, rebounds off the leading shock and mixes back in with older lobe material (i.e. it continues to do what the jet was doing before switch off). For large sources, the bulk flow continues in the direction of the declining ambient profile. Our 20 kpc and 60 kpc switch-off simulations are confined to the inner 70-50 kpc for most of their active and remnant life. In this region, the ambient density only drops by a factor of two and the preferential direction of fluid flow is not along the jet axis. The evolution of these sources is then dependent on the magnitude of the velocities within the remnant lobe and the inhomogeneities in the environment. As shown in the right-hand panels in the bottom three rows of Fig \ref{fig:remnants}, this results in a greater loss of structure at $t_{\rm{on}} + 50$ Myr. We note that the simulations presented in this paper do not contain magneto-hydrodynamic effects, which may suppress instabilities at the boundary between the lobe and shocked shell \citep{Shabala_Alexander_feedback_2009}. However, we expect the modelling of flows within the lobes to be representative as these flows are governed by pressure gradients, and large-scale hydrodynamic interactions.

\subsection{Lobe Dynamics}
\label{subsec:lobe_dynamics}
We now probe the dynamic evolution of the active and remnant lobes in more detail. We explore the evolution of models with low kinetic power progenitors ($Q_{\rm{j}} = 10^{36}$ W) in Section \ref{subsec:lpp-lobe_dynamics} and those with high-power progenitors ($Q_{\rm{j}} = 10^{38}$ W) in Section \ref{subsec:hpp-lobe_dynamics}. 

\subsubsection{Low-Power Progenitors}
\label{subsec:lpp-lobe_dynamics}

\begin{figure*}
    \centering
    \includegraphics[width=0.7\linewidth]{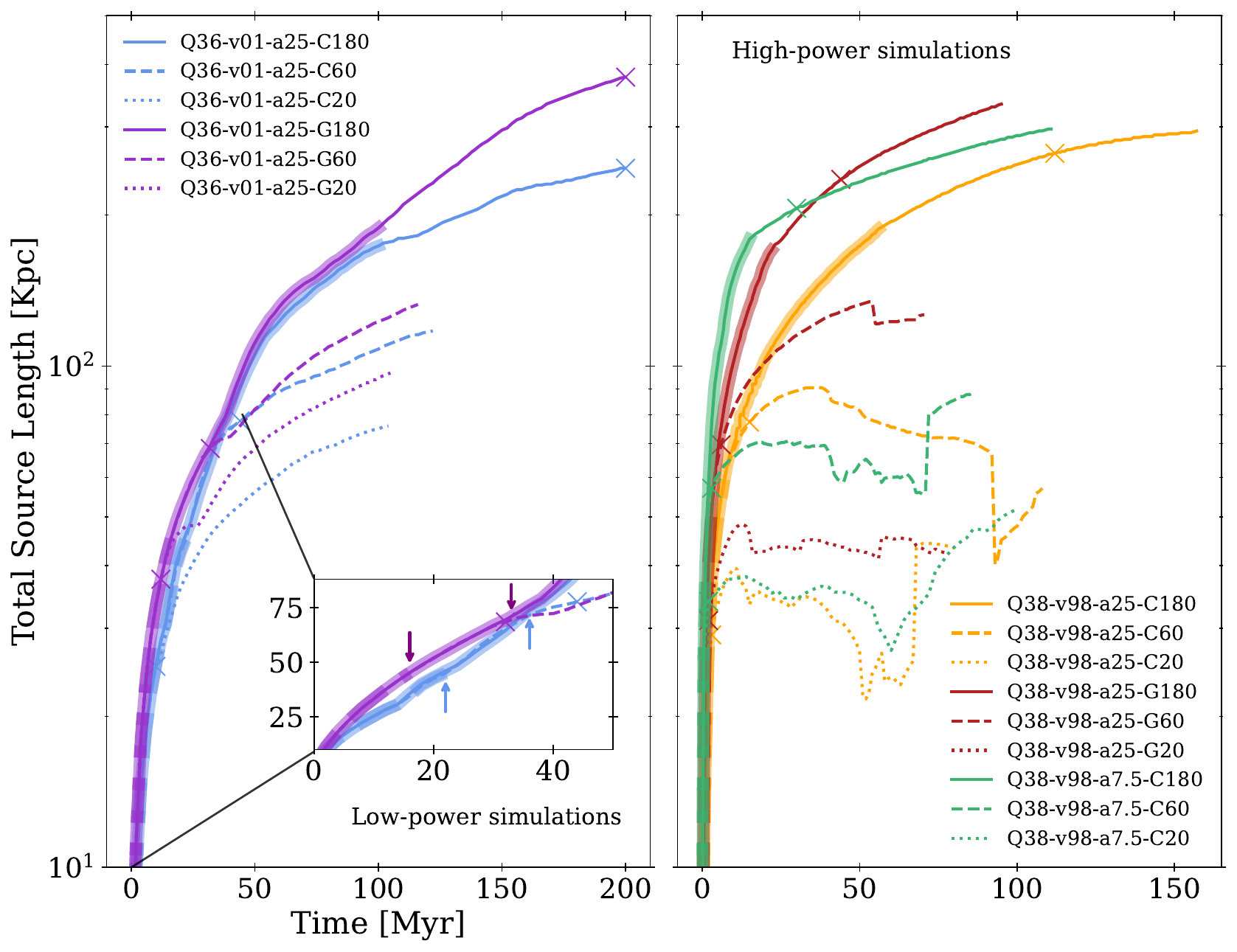}
    \caption{The time evolution of the total source length measured as the maximum extent (along $z$) of the tracer material above $10^{-4}$. The left-hand panel shows all models with low injected jet kinetic powers ($Q_{\rm{j}} = 10^{36}$ W). Models with high injected jet kinetic powers ($Q_{\rm{j}} = 10^{38}$ W) are shown on the right. The active and remnant phases are shown with thick and thin lines, respectively. The cross-marks indicate $2\times t_{\rm{on}}$. The inset figure shows a zoomed-in view of the 60kpc switch-off low-powered simulations from 0 to 50 Myr showing how these small sources continue to grow as if they were active for some time. The small arrows show the time delay between the point where the jet flow stops and the point where the lobe length evolution begins to slow down.}
    \label{fig:length_evolution}
\end{figure*}

The left-hand panel of Fig. \ref{fig:length_evolution}, shows the time evolution of the total source length along the jet propagation axis. This is measured from the maximum extent along the jet axis of the tracer fluid above $10^{-4}$. For $t < t_{\rm{on}}$, a key feature of our low-power sources is that equivalent jet models in different environments grow at a similar rate and have similar total active times for a given size. This is shown by the proximity and slope of the thick blue and purple lines in Fig. \ref{fig:length_evolution}. This is not an intuitive result. Since the central density and pressure in the group environment are approximately an order of magnitude below the cluster, one would expect all models in the group to propagate faster. However, in the group simulations, the larger jet radius at collimation (discussed in Section \ref{subsubsec:active_source_properties}) and the greater lateral expansion of the lobe plasma cause the momentum flux to dissipate over a larger area. \\

For low-power 20 kpc and 60 kpc switch-off simulations (see the left panel of Fig. \ref{fig:length_evolution}) there is a time delay between the jet switch off and the divergence of the length evolution from the trajectory of the active source. This is due to the travel time of the trailing jet material through the lobe. To see this more clearly, we isolate the 60 kpc switch-off simulations in the inset in Fig. \ref{fig:length_evolution}. Here, the arrows show the difference in time between the jet switch off and the point where the length evolution starts to separate from the active track. Hence, the smaller 20 and 60 kpc switch-off simulations continue to expand as if they were still active for almost 20 Myr. For the largest group simulation (Q36-v01-a25-G180; purple solid lines), we see an increase in total source expansion when the remnant phase begins. We attribute this to the low ambient density and pressure in this region of the environment (see Fig. \ref{fig:environment comparison}).\\

\begin{figure*}
    \centering
    \includegraphics[width=0.7\linewidth]{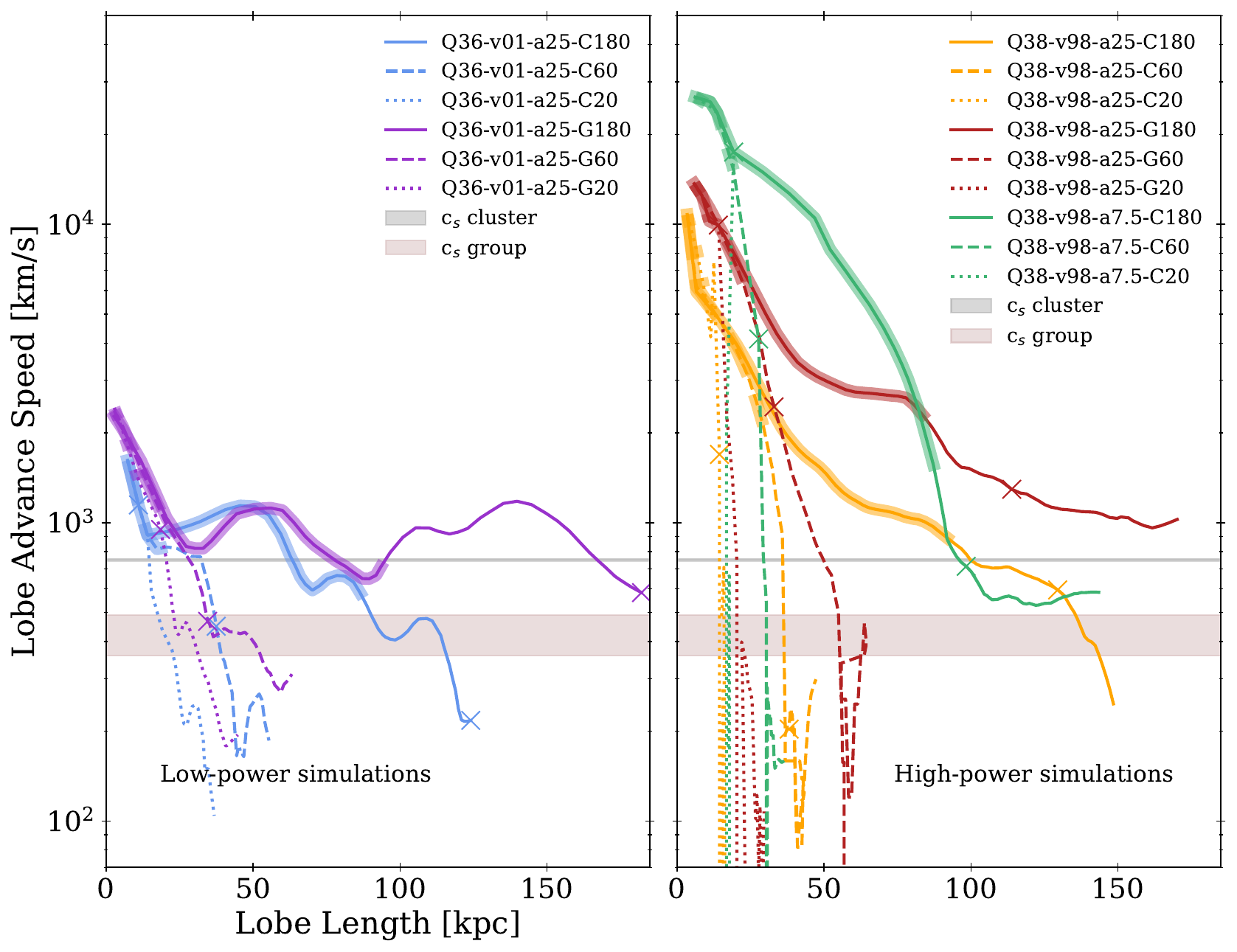}
    \caption{The forward (propagating in the +z direction) advance speed of the primary lobe shown against lobe length, illustrating the deceleration of all sources after switch-off. The figure features are consistent with Fig. \ref{fig:length_evolution} where the left-hand panel shows all models with low injected jet kinetic powers and high powers are on the right and the thick to thin lines denote the active and remnant phases, respectively. The cross marks indicate 2$\times t_{\rm{on}}$.}
    \label{fig:advance_velocities}
\end{figure*}

To determine when the expansion velocity of the sources matches the average sound speed in the ambient medium, we show the time derivative of the source length evolution (Fig. \ref{fig:length_evolution}) in Fig \ref{fig:advance_velocities}. To reduce the noise, we smooth the data using a LOcally WEighted Scatter plot Smoothing model (LOWESS\footnote{we use the LOWESS model implemented in the Python package, \texttt{statsmodels} \citep{statsmodels}. A bandwidth of 0.2 is used such that the final smoothed model follows the original data closely.}). The range of sound speeds in the undisturbed cluster and group environments are shown by the grey and red shaded regions, respectively. In the cluster environment, before $t_{\rm{on}}$, the expansion rates are transonic/subsonic compared to local sound speeds. In the group, our 20 kpc and 60 kpc switch-off sources eventually transition to subsonic velocities. We note here that these sources  are strongly affected by the dynamics of the environment, making estimates of their forward velocity difficult after switch off.  Simulation Q36-v01-a25-G180 continues to expand at least mildly supersonically (a Mach number between 1-2) for the duration of the simulated time.\\

\begin{figure*}
    \centering
    \includegraphics[width=0.7\linewidth]{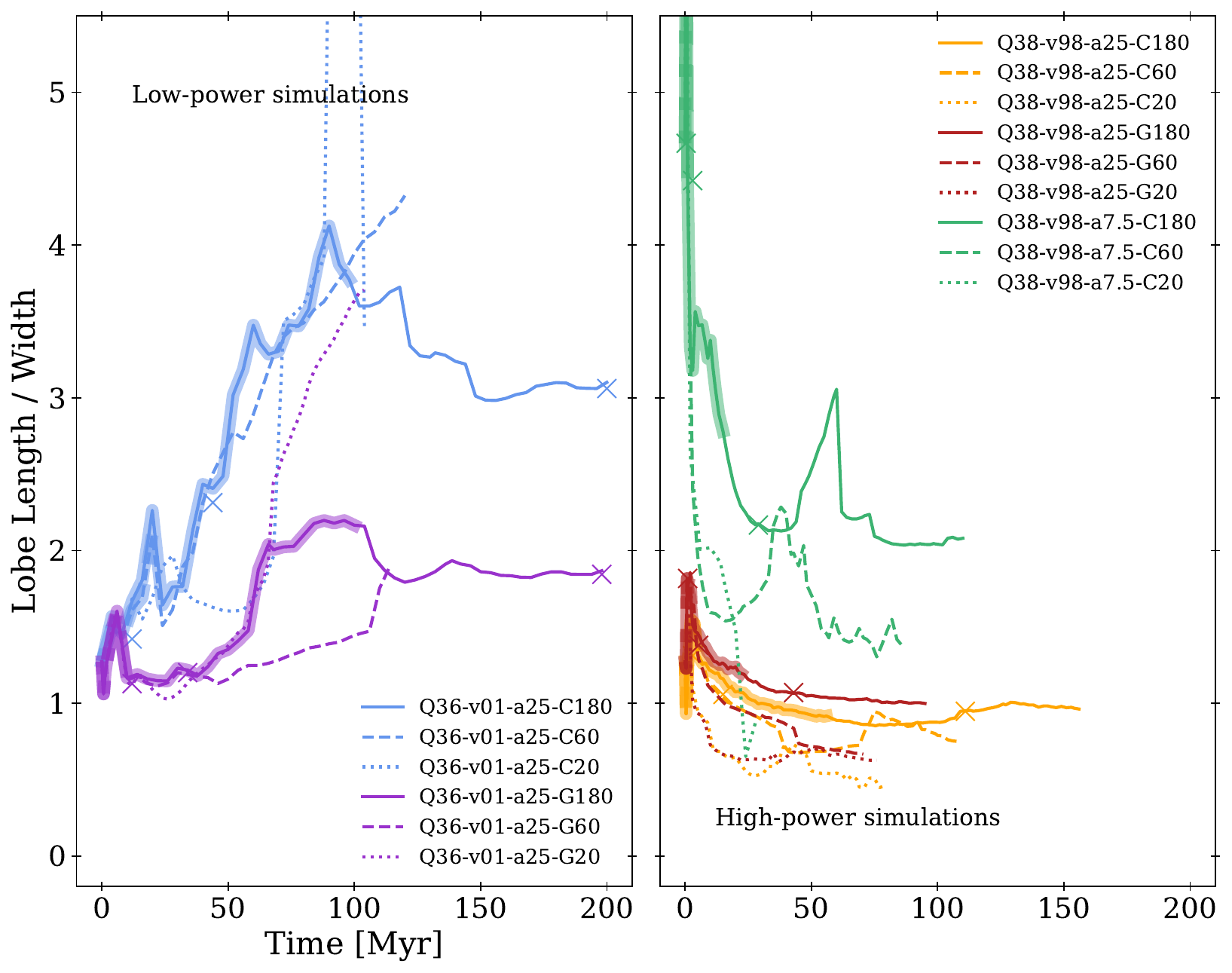}
\caption{The time evolution of the ratio of the primary lobe length to width. The lobe width is measured as the maximum $\pm x$ extent in a narrow slice centred at the lobe midpoint.}
    \label{fig:axial_ratios}
\end{figure*}

We track the evolution of the length to width ratio of the primary lobe in Fig. \ref{fig:axial_ratios}. The length to width ratio is defined as the ratio of the full length of the lobe to the maximum width measured at the lobe midpoint. We consider only the lobe propagating in the $+z$ direction, though results are similar for the $-z$ lobe. With this definition, low length to width ratios imply wide lobes, while high length to width ratios imply narrow ones. The initial spike in the length to width ratio tracks at $t\sim5$ Myr corresponds to the jet punching through the central region of the host environment before the lobe has inflated. This is similar to the jet-breakout phase described in \cite{Sutherland_2007} and  \cite{Turner_etal_2023}.\\

The switch-off time of the jets has some effect on the length to width ratio over the entire lifetime of the low-power simulations. In large, 180 kpc switch-off models, the length to width ratio stabilises, suggesting quasi-self-similar expansion after switch-off. Conversely, in 20 kpc and 60 kpc switch-off simulations, the length to width ratio increases as trailing jet material burrows through the top of the lobe and rises away from lower radii plasma. The environment also plays a role; the higher density and pressure in the cluster environment results in narrower lobes with consistently higher length to width ratios. This occurs because the surrounding pressure in the cluster environment suppresses lateral expansion. The forward (jet-driven) expansion is less affected, allowing the jets to drill through the ambient medium.\\

\begin{figure*}
    \centering
    \includegraphics[width=0.78\linewidth]{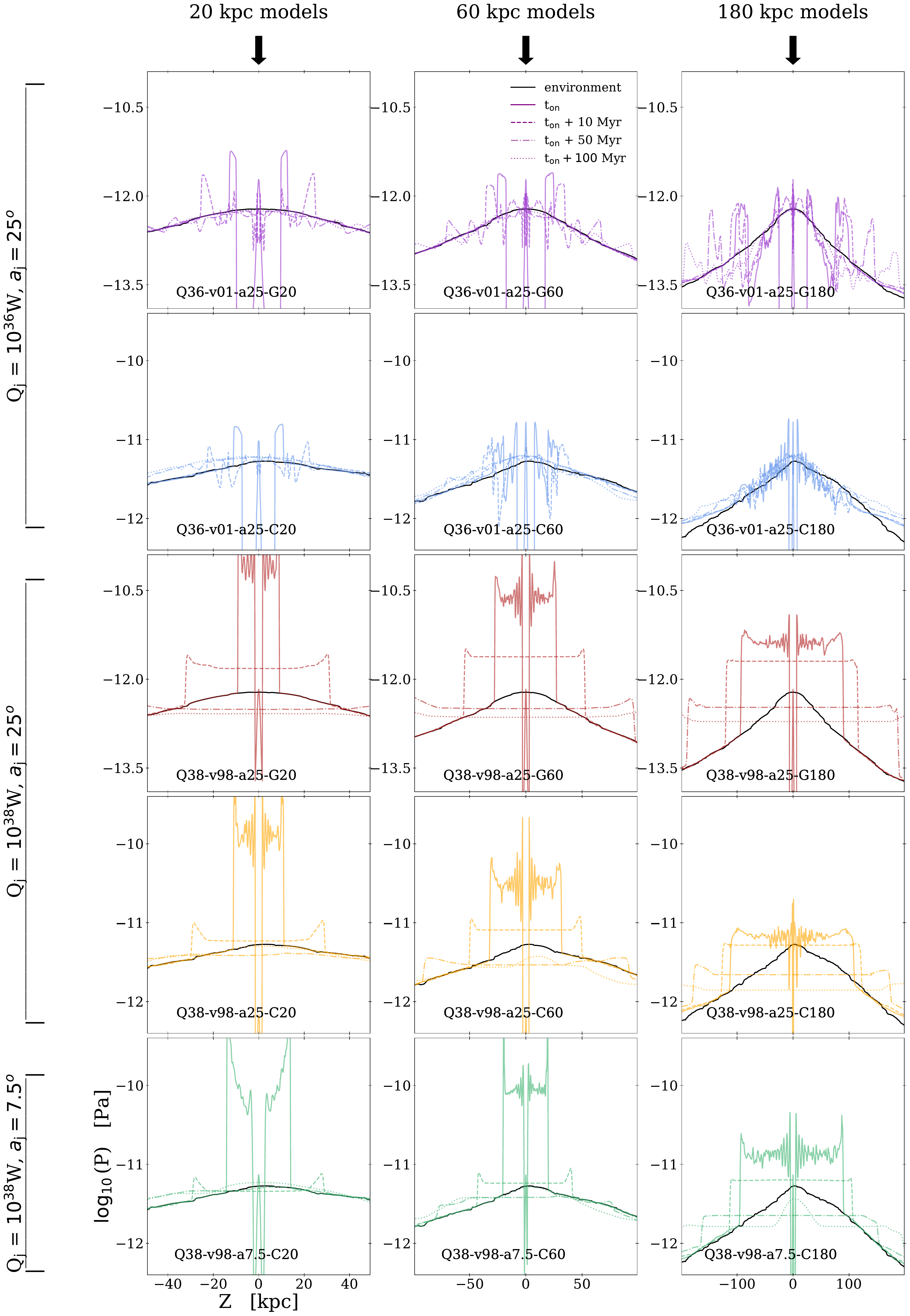}
    \caption{Presssure profiles along the axis of jet propagation for all simulations showing how remnants of low-power (top two rows) and high-power (bottom three rows) progenitors affect the initial gas distribution (solid black line). We show the pressure distributions at $t_{on}$ (solid colored line) and then 10 Myr (dashed), 50 Myr (dot-dashed), and the maximum simulated time (indicated in Table \ref{tab:sim_props}, dotted line) in the remnant phase. }
    \label{fig:pressures}
\end{figure*}

We lastly consider the pressure profiles of all sources to probe the effect the active and remnant evolution have on the initial gas distribution. We show pressure profiles along the axis of jet propagation for all sources in Fig. \ref{fig:pressures}. In the active phase, the pressure profiles of low-powered sources are characteristically non-uniform, with regions of underpressure and slight overpressure. The regions of underpressure are located in the wake of the expanding head of the lobe and are more than an order of magnitude less than the pressure of the initial ambient profile. In the remnant phase, all low-powered simulations quickly return to pressure balance with the initial profile.

\subsubsection{High-Power Progenitors}
\label{subsec:hpp-lobe_dynamics}

We now consider the evolution of our high-power ($Q_{\rm{j}} = 10^{38}$ W), fast ($v_{\rm{j}} = 0.98$c) sources using the same methods as for the low-power sources. During the active phase, all high-power fast sources are significantly overpressured and expand supersonically through the ambient medium. The right-hand panels of Fig. \ref{fig:length_evolution} and \ref{fig:advance_velocities} show that simulations expanding into the group have faster rates of expansion and take roughly half the time to reach the same length; in contrast to low-power sources, where group simulations were only slightly faster and the active times for equivalent jet simulations were similar. As the lobes inflate, the length to width ratio of all high-power models decreases (see Fig. \ref{fig:axial_ratios}) but will stabilise in most cases within a few Myr after switch-off, indicating quasi-self similar expansion. At later times in the remnant phase, the remnant lobe structure can become susceptible to instability growth and the weak motions of the ambient medium which cause fluctuations in the length to width ratio tracks. Since our measurement of length to width ratio takes the total lobe width at the lobe midpoint, spikes in the length to width ratio tracks of Q38-v98-a25-C180 (green lines) are reflective of `pinching' that starts to develop in the main body of the lobe during the remnant phase. The onset of this `pinching' can be seen in the bottom right panel of Fig. \ref{fig:remnants}. \\

The deceleration of the jet plasma and subsequent transition to a state of pressure equilibrium is set by the relative balance between jet-dominated and lobe-dominated expansion. Both wide and narrow opening angle simulations are jet-dominated within the first few Myr (as indicated by higher length to width ratios during this time). For narrow opening angles, the expansion continues to be more jet-dominated out to larger source sizes, with mildly relativistic jet velocities maintained out to approximately 90\% of the lobe length. Jet-dominated simulations decelerate most rapidly once the jet flow ceases; as seen in the right-hand panel of Fig. \ref{fig:advance_velocities} however, this deceleration is also seen prior to switch-off. For wide opening angles, the transition from a jet-dominated to lobe-dominated expansion occurs quickly after the lobes begin to form. \\

As shown in the bottom three rows of Fig. \ref{fig:pressures}, all high-powered remnants are strongly overpressured in the active phase and display a uniform pressure distribution along the symmetry axis, as expected from their combination of injection power and velocity. In the remnant phase, the lobes remain overpressured (relative to the material outside the lobe/shocked ambient system) but the pressure distribution in the central region of the grid falls below the initial profile as the lobes continue to drive gas from low radii. In one simulation (Q38-v98-a7.5-C180 in the lower right of Fig. \ref{fig:pressures}), we see an increase in pressure of the central region at late times (the dotted green line) as gas refills the central region and pushes aside the remnant lobes.

\subsection{Bow Shock Dynamics}
\label{subsec:shocks}

We now consider the evolution of the surrounding bow shock. Initially, all simulations in this work expand supersonically, driving shocks into the ambient medium. These regions can be seen in Fig. \ref{fig:actives} and \ref{fig:remnants} as a parabolic-shaped cross-section of hot, high-density fluid surrounding each radio jet and jet-inflated lobe. We track the leading edge of the shocked region by identifying the point where the pressure gradient is maximised outside the lobe material. This matches well with the visible trace of the shocked region seen in Figs. \ref{fig:actives} and \ref{fig:remnants}. We track the bow shock until the maximum simulation time is reached, the pressure gradient is less than $1.1$ per grid cell (a necessarily low threshold to capture weak shocks while still avoiding picking up fluctuations in the ambient pressure profile), or it has exceeded the grid boundary. Low powered sources are discussed first in Section \ref{subsec:lpp-shock_dynamics} and then high-power sources in Section \ref{subsec:hpp-shock_dynamics}.

\begin{figure*}
    \centering
    \includegraphics[width=\linewidth]{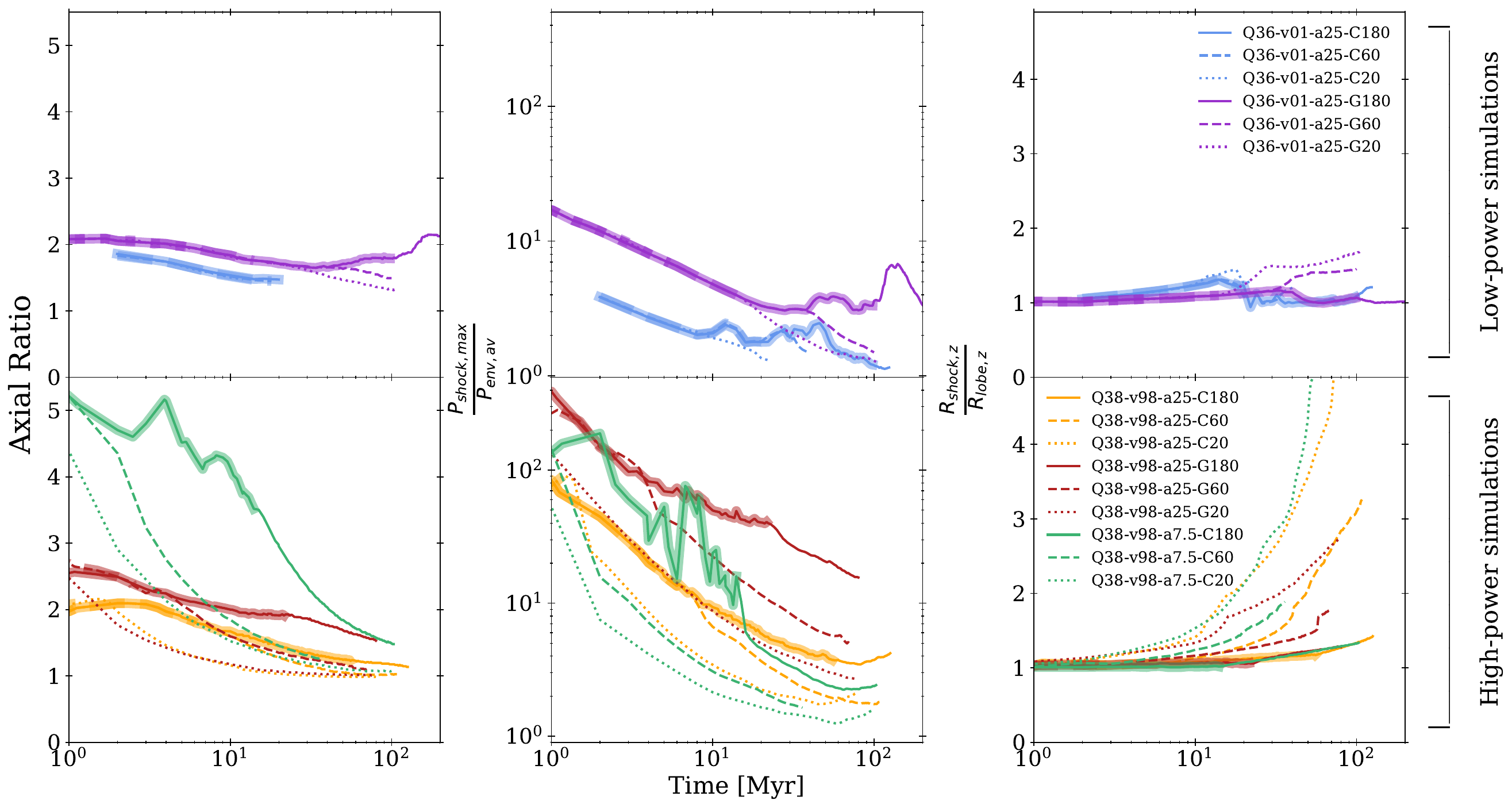}
    \caption{Evolution of the length to width ratio (left) of the shocked region, the ratio of the maximum shock front pressure to the undisturbed ambient medium (middle), the ratio of shock radius to lobe radius along the $z$ direction (right). As before, thick lines denote active sources and thin ones denote remnants. Tracks of low-power sources are given in the top row and high-power sources are given in the bottom row. For the low-power cluster simulations, the strength of the lateral shock becomes indistinguishable from the ambient medium at around $20$ Myr and so the blue lines in the right-hand panel are truncated. \\}
    \label{fig:shock_properties}
\end{figure*}

\subsubsection{Low-Power Progenitors}
\label{subsec:lpp-shock_dynamics}

All low-power simulations produce weak bow shocks that do not considerably change in shape or separate from the lobe during the remnant phase. In the left-hand panel of Fig. \ref{fig:shock_properties}, we show the length to width ratio of the shocked region which we measure as the ratio of the forward shock radius to the lateral radius taken along the positive x-axis. Very similar results are produced if the y-axis is considered due to the spherical symmetry of the shock. For both group and cluster simulations, the length to width ratio values remain in the range of $1.5-2.2$ over the entire simulated time and the shock fronts separate only marginally from the head of the lobe as shown in the right panel of Fig. \ref{fig:shock_properties}. For low-powered sources, the shock separation is more pronounced for small remnants in the group environment, likely as a result of the low-density declining atmosphere.\\

The strength of the shock front is presented in the middle panel of Fig. \ref{fig:shock_properties}. At the start of the remnant phase, the pressure of the shock front for the low-power cluster simulations is, at most, twice that of the ambient medium. As a result, we are only able to track the shock front for about $20$ Myr into the remnant phase before it drops below $1.1\times P_{\rm{env}}$. For the group simulations, the shock pressure ratios are higher at the end of the active phase (around $3\times P_{\rm{env}}$) but take over six times longer to decay to similar values as the ambient medium.

\subsubsection{High-Power Progenitors}
\label{subsec:hpp-shock_dynamics}

Our high-power, fast simulations are characterised by strong bow shocks, and they can be distinguished from the ambient medium for the total simulated time. Before the remnant phase begins, the bow shock is between $3-200$ times more overpressured than the undisturbed ambient medium (see the lower middle panel of Fig. \ref{fig:shock_properties}) and is only marginally separated from the cocoon at the head of the lobe (lower right panel of Fig. \ref{fig:shock_properties}). When the remnant phase starts, the bow shock progressively separates from the head of the lobe and tends towards sphericity as shown by the decreasing length to width ratio values in the bottom left panel of Fig. \ref{fig:shock_properties}. This result is consistent with a decrease in jet head expansion velocities. A similar observation is made by \cite{Perucho_2011} and \cite{Perucho_2014} for their suite of strongly relativistic jets with kinetic powers in the range of $10^{37} - 10^{39}$ W. \\

The separation between lobe and bow shock is most drastic in the smallest switch-off simulations, where the edge of the shocked region can increase to twice the lobe radius within 20 Myr. Meanwhile, the bow shock pressure ratio falls to below 20. The initial drop in pressure at the start of the remnant phase is most rapid for narrow opening angle simulations (the solid green track in Fig. \ref{fig:shock_properties}). For high-power, wide opening angle simulations in the group environment, the rate of bow shock pressure decrease is similar the equivalent cluster model, but it remains more than $10$ times overpressured relative to the surrounding environment. 


\section{DISCUSSION}
\label{sec:discussion}

In this section, we compare our results to existing numerical and analytical modelling of remnant radio sources in Section \ref{subsec:cocoon and shock evolution}. In Section \ref{subsec:onbuoyancy} we discuss the relative role of jet momentum flux and buoyancy in driving the expansion of our sources in the active and remnant phases and show that simple analytic expectations can provide good estimations of the onset of different evolutionary phases. In Section \ref{subsec:remnant_heaters?}, we consider whether remnants can continue to provide feedback on the ambient medium after switch-off.

\subsection{Benchmarking}
\label{subsec:cocoon and shock evolution}

\subsubsection{Numerical Modelling}
\label{subsubsec:discussion-numerical_modelling}

We start by comparing our results to those found in similar, past works. The evolution of large-scale (several hundred kpc), high-powered ($>10^{37}$ W) sources has been well studied by authors including \cite{Walg_2014}, \cite{Perucho_2014} and \cite{English_2019}. The evolution of the high-powered sources in the present study produces similar key results and features. Similar to \cite{Perucho_2014} and \cite{English_2019} we show that the shocks and lobes tend towards sphericity once the driving forward momentum is removed. As the lobe material decelerates post switch-off, the shock front separates and weakens as it propagates into the ambient environment (see Fig \ref{fig:shock_properties}). As expected and in line with the results of \cite{Walg_2014}, the trailing jet material disappears on timescales set by the length of the cocoon and the bulk velocity of the jet material. This is also the duration of time that the source will continue to `masquerade' as if it were still active. For our high-powered sources, the jet material traverses the lobe in a shorter time than typical time between simulation outputs. Hence, these sources appear to deviate from their active behaviour immediately following switch-off.\\

We find the environment to be important for the dynamics of the simulated sources regardless of the jet kinetic power and injection speed. A key difference is the amount of overpressure in the lobes. For large-scale high-power sources in group environments, the source can remain significantly overpressured in the outer parts of the lobe for much longer than the previous active phase, as shown in Fig. \ref{fig:pressures}. This is largely attributed to the lower ambient pressure in this environment. The same behaviour is seen for low-power sources. Although the overpressure factor is lower for low-powered sources, the volume occupied by the lobes is a factor $2-3$ larger in the group than in the cluster at a given source length. In \cite{English_2019}, the authors find their remnant lobe dynamics to be less dependent on the environment, with all models continuing to expand at similar rates after switch-off. We show that while equivalent high-power group and cluster models decelerate at a similar rate initially, the difference in environment properties results in the lobe expansion speed of group simulations reaching the ambient sound speed later than their cluster counterparts. We further explore the transition to buoyancy in Section \ref{subsec:onbuoyancy}.

\subsubsection{Analytic Modelling}
\label{subsubsec:discussion-analytic_modelling}

We now look at how our simulation results compare with analytic modelling and investigate whether existing analytic models can be used to capture the evolutionary histories of simulated remnant radio galaxies. The transition of a powerful, relativistic source from its active to remnant phase has been explicitly considered in the literature with different approximations. \cite{Krause_2005} derived solutions for the shock radius in a spherical thin-shell model, including the gravity of the dark matter halo. A self-similar model without gravity, but for both lobe and shock radius, was derived by \cite{Kaiser_Cotter_2002}. In that work, the authors assume a spherical radio source with uniform internal pressure and derive a set of steady-state similarity solutions that describe the evolution of the cocoon and shock radii in the active and remnant phases. In the active phase, the radio source evolution can be described by their Eq. 5. Once the jet switches off, \cite{Kaiser_Cotter_2002} assume that the cocoon remains overpressured and immediately enters a coasting phase. For the case where the adiabatic index of the cocoon and shocked material is $5/3$ representing non-relativistic material (a reasonable assumption for the material in the lobes of our simulations), the growth of the cocoon is given by $R \propto t^{\alpha}$ where, for the cocoon, $\alpha_{\rm{c}} = \frac{2(\Gamma + 1)} {\Gamma(7+3\Gamma - 2\beta)} = \frac{16}{5(12-2\beta)}$, and for the shocked shell, $\alpha_{\rm{s}} = \frac{4}{7+3\Gamma-2\beta} = \frac{4}{12-2\beta}$ where $\beta$ is the log slope of the ambient density profile $\rho \propto r^{-\beta}$.\\

The steady-state solutions in \cite{Kaiser_Cotter_2002} ignore the lobe energetics during the transition between the active and remnant states. If the system takes time to settle into a coasting phase, then it will not be well described by the \cite{Kaiser_Cotter_2002} solution. We follow the work presented in \cite{Turner_etal_2023} and \cite{Turner_Shabala_2023}, and derive a set of coupled differential equations that describe the expansion of the cocoon and shocked region. These are obtained from Eqs. 1, 2 and 3 in \cite{Kaiser_Cotter_2002} but are recast such that the current expansion rate is influenced by the source's expansion history. We assume that the radio source is elongated along the jet axis with an axis ratio, $A$. In this case, $A$ is defined as the length of the lobe divided by its maximum transverse radius (not full lobe width), as given in \cite{Turner_etal_2023}. The values for $A$ that we derive from our simulations are then approximately twice those in Fig. \ref{fig:axial_ratios} at the switch-off point. Following \cite{Turner_etal_2023}, we implement a standard fourth-order Runge-Kutta method which solves the following equations for cocoon and shock radii:

\begin{equation}
\begin{split}
    \dot{R_{\rm{c}}} & = \frac{(\Gamma - 1)Q_{\rm{j}}A^2 R_{\rm{s}}^\beta}{2\pi\Gamma R_{\rm{c}}^2k\dot{R_{\rm{s}}}^2} 
    + \frac{\beta R_{\rm{c}} \dot{R_{\rm{s}}}}{3R_{\rm{s}}\Gamma} 
    - \frac{2R_{\rm{c}} \ddot{R_{\rm{s}}}}{3\Gamma\dot{R_{\rm{s}}}} \\
    \ddot{R_{\rm{s}}} & = \frac{3(\Gamma-1)Q_{\rm{j}} R_{\rm{s}}^{\beta-3}A^2}{4\pi k \dot{R_{\rm{s}}}} 
    + \frac{\dot{R_{\rm{s}}}^2(2\beta-3\Gamma-3)}{4R_{\rm{s}}},
\end{split}
\end{equation}

\noindent where $k = \rho_{\rm{0}} a_{\rm{0}}^\beta$ in which  $a_0$ and $\beta$ are, respectively, the core radius and exponent in a simple power-law description of an ambient gas density profile, $\rho(r) = \rho_0(r/a_0)^{-\beta}$. Since our cosmological environments are a poor match for the simpler power-law prescription used by \cite{Kaiser_Cotter_2002}, we use the approach of \cite{Turner_2015} and approximate the simulated gas density profile as being made up of several sequential, contiguous power-laws. The $\beta$ for each is found by taking the density slope in the region of interest. \\
  
\begin{figure}
    \centering
    \includegraphics[width=\linewidth]{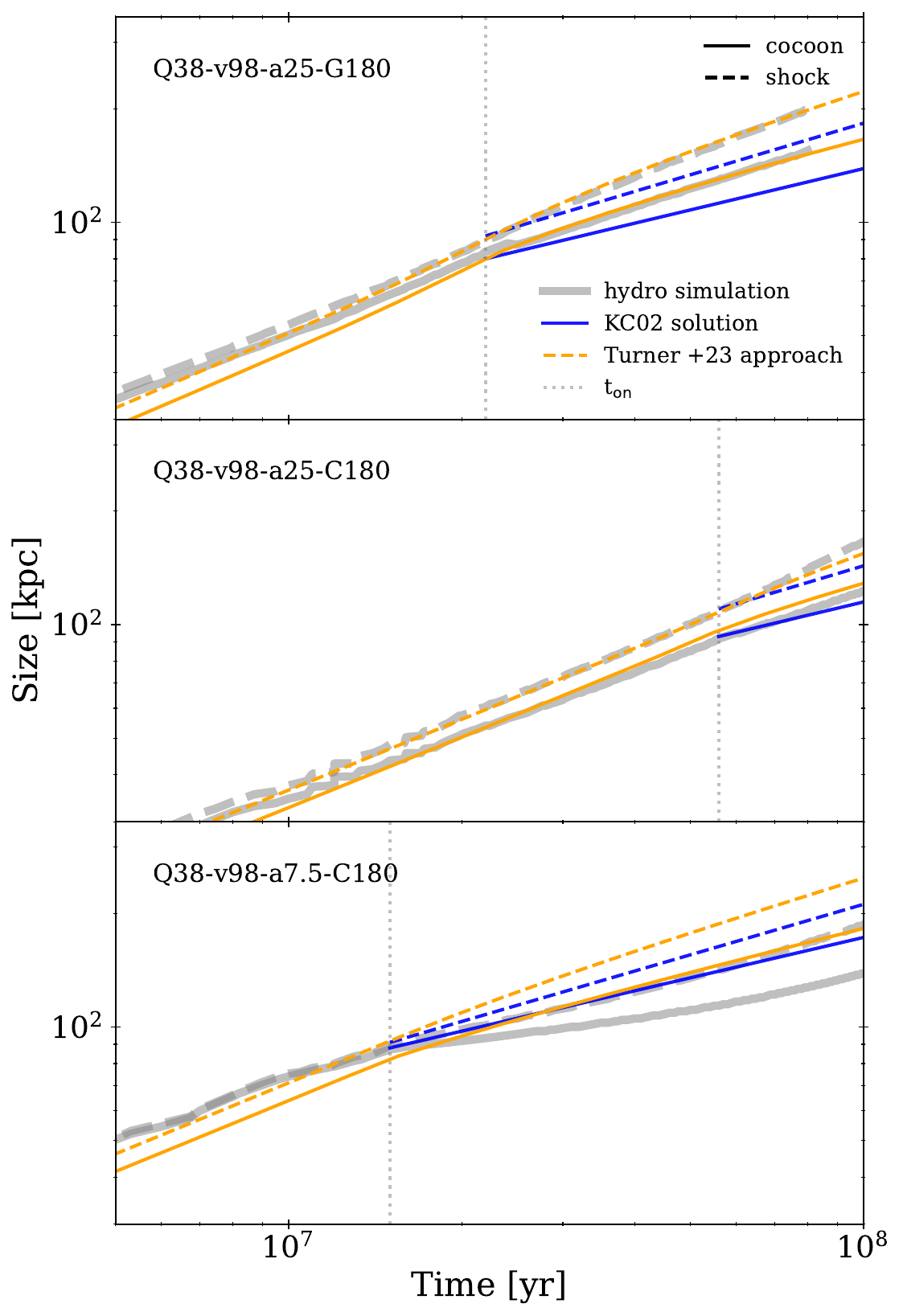}
    \caption{A comparison of the results of the shock (dashed lines) and cocoon (solid lines) evolution for three high-power simulations. The orange and blue sets of lines respectively show the expected shock and cocoon evolution for our differential approach and the analytic solution of \cite{Kaiser_Cotter_2002}. The point at which the simulation becomes a remnant is indicated by the vertical grey dotted line.}
    \label{fig:KC02_comparison_1}
\end{figure}

We show a comparison between a subset of our numerical simulations, the expectations of the \cite{Turner_etal_2023} approach, and the expectations of the \cite{Kaiser_Cotter_2002} analytic equations in Fig. \ref{fig:KC02_comparison_1}. The key result is that, for sources whose active expansion is not dominated by the jet momentum thrust (e.g. our high-power, wide opening angle models), our analytical approach (orange lines) shows close agreement with the numerical simulations (grey lines). However, where the expansion is jet-dominated, both analytic approaches overestimate the source size into the remnant phase because they do not take the initial jet expansion phase into account.

\subsection{Buoyancy}
\label{subsec:onbuoyancy}

Observations from \cite{Brienza_2017}, \cite{Mahatma_2018} and \cite{Jurlin_2021} show that candidate remnants can have amorphous morphologies at
MHz frequencies. In fact, this type of morphology is often taken as an indicator of remnant status \citep{Brienza_2016}. Dynamically, these relaxed structures would be expected to develop following the onset of buoyancy \citep[e.g.][]{Churazov_2001}. In this section, our goal is to determine when this takes place from our numerical results, and provide an analytical description of the buoyant regime in both the active and remnant phase. We analyse the active phase in Section \ref{subsubsec:discussion-buoyancy-active_sources} and the remnant phase in \ref{subsubsec:discussion-buoyancy-remnant_sources}.

\subsubsection{Active Sources}
\label{subsubsec:discussion-buoyancy-active_sources}

We first must determine whether our simulations are buoyancy dominated before the onset of the remnant phase. \cite{Komissarov_1998} and \cite{Krause_2012} describe a length scale, $L_{\rm{2}}$, where a radio source of kinetic power $Q_{\rm{j}}$ would be expected to come into pressure equilibrium with the surrounding environment,

\begin{equation}\label{eq.L2on}
    L_2 = \left(\frac{Q_{\rm{j}}}{\rho_{\rm{ext}} c_{\rm{s, ext}}^3}\right)^{1/2}
\end{equation}

\noindent where the external sound speed, $c_{\rm{s, ext}} = \left(\frac{\Gamma_{\rm{ext}} P_{\rm{ext}}}{\rho_{\rm{ext}}}\right)^{1/2}$,  is calculated from the local  ambient pressures $P_{\rm{ext}}$ , and densities, $\rho_{\rm{ext}}$ , through which the radio source propagates. We reasonably assume that the ambient medium is well-described by an ideal adiabatic index of $\Gamma$ = 5/3. For sources with linear sizes $L < L_2$, the lobe plasma is expected to be supersonic, overpressured and the forward expansion is driven by ram pressure. \\

The local ambient sound speeds range from around $700 - 780$ km/s in the cluster and $360-480$ km/s in the group. Fixing the local sound speed to a median value yields $L_2 = 16$ kpc in the cluster and $L_2 = 70$ kpc in the group for low-power sources. The same exercise repeated for high-power sources gives $L_2 = 160$ kpc in the cluster and 700 kpc in the group (well beyond the confines of the grid). For all low-power sources in the cluster environment, the switch-off length is in excess of the value of $L_2$, hence we expect these sources to be strongly buoyancy-driven prior to the start of the remnant phase. We now consider how the expected terminal rise speeds of a buoyant bubble compare to the local sound speed.\\

The terminal rise velocity of a buoyant bubble can be estimated following the work of \cite{Churazov_2001}, \cite{Ensslin_2002}, and \cite{Perucho_2014} under the assumption that the density of the bubble is sufficiently below the density of the ambient environment. We equate the buoyant force experienced by a bubble of cross-sectional radius, $R_{\rm{b}}$ in an environment defined by gravity $g$ and density $\rho_{\rm{ext}}$, as $F_{\rm{buoy}} = \frac{4}{3}\pi R_{\rm{b}}^3g\rho_{\rm{ext}}$, with the drag force provided by the ram pressure as the plasma moves through the external environment $F_{\rm{drag}} \sim C_{\rm{d}}v_{\rm{b}}^2\rho_{\rm{ext}} \pi R_{\rm{b}}^2$. The cosmological environments used in this work were selected to be as close as possible to hydrostatic equilibrium (see \PaperI). In such environments, the gravitational effect, $\rho_{\rm{ext}}(r)g$, can be approximated as the pressure gradient in the considered region $\frac{dP_{\rm{ext}}}{dr} \simeq -\rho_{\rm{ext}}(r)g$. The bubble velocity is then given by 
\begin{equation}\label{eq.bubblerisevelocity}
    v_{\rm{buoy}}(r)^2 = \frac{4R_{\rm{b}} }{3C_{\rm{d}}\rho_{\rm{ext}}(r)} \frac{dP_{\rm{ext}}}{dr}
\end{equation}

Similar to \cite{Churazov_2001} we assume a drag coefficient, $C_{\rm{d}} \approx 0.75$. For a bubble of radius 3-10 kpc (the radius of our smallest sources in the active phase) moving through the inner 50 kpc of our cluster environment, the terminal rise speeds are expected to be half of the local sound speed. We expect that these sources are still significantly driven by pressure during the active phase. For larger sources, the terminal rise velocity in the region $50-90$ kpc from the environment core is comparable to the sound speed. Based on our estimate of $L_2$ and the expected rise velocities of sources in this region, we do not expect any high-power simulations to be buoyant before the remnant phase.

\subsubsection{Remnant Sources}
\label{subsubsec:discussion-buoyancy-remnant_sources}

\begin{figure}
    \centering
    \includegraphics[width=0.95\linewidth]{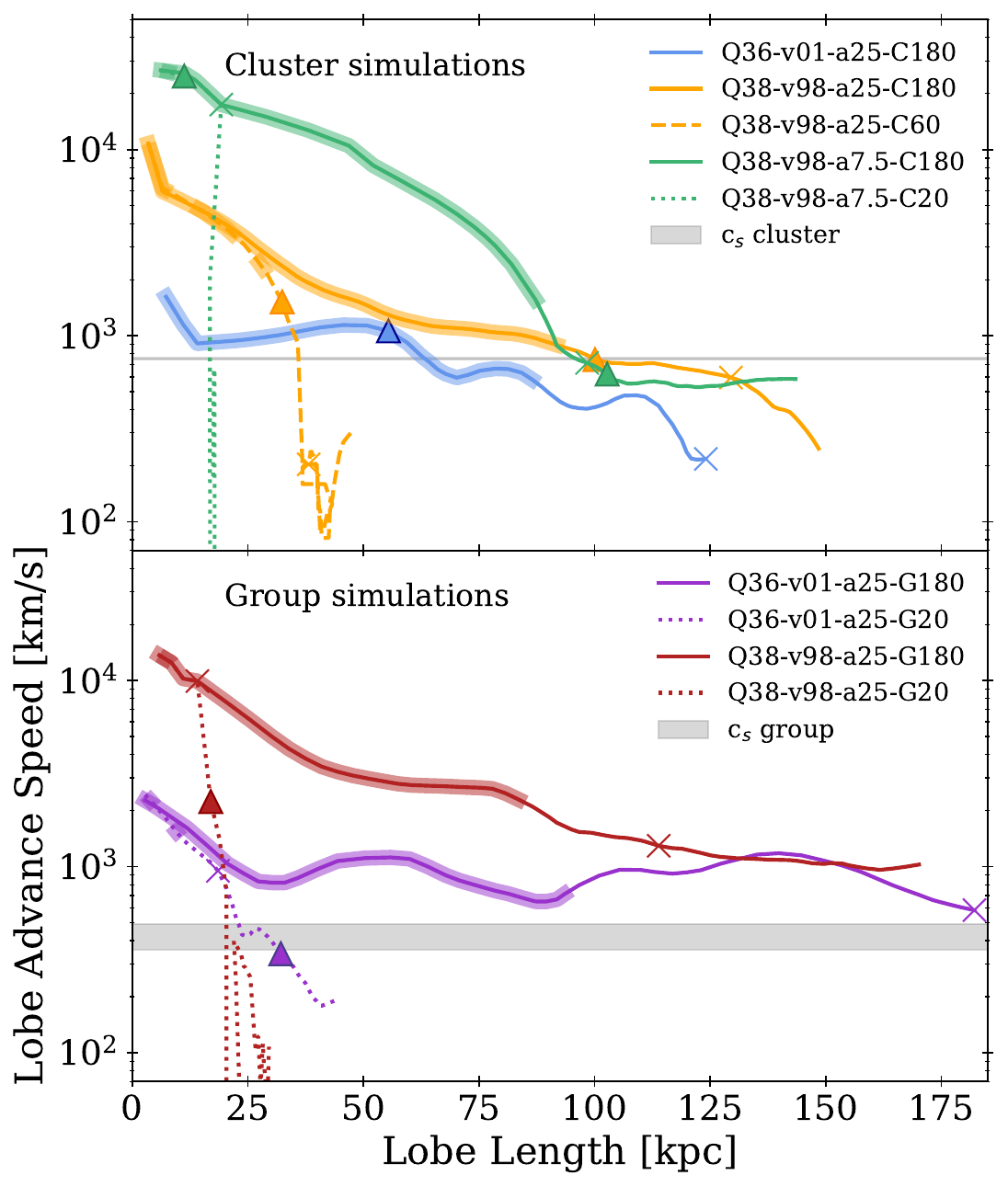}
    \caption{Forward advance velocities for a subset of our simulations showing the active (thick) and remnant (thin line) phases for these simulations. The predictions of the buoyant, remnant length scale $L_{2,\rm{rem}}$ are denoted by the triangle markers. The top panel shows cluster simulations and the bottom panel shows group simulations. For the large group simulations shown in the lower panel, the remnant length scale is larger than the grid size and hence, triangle markers are not shown. The cross marks indicate $2t_{\rm{on}}$ and the grey shaded regions show the range of sound speeds in the ambient medium.}
    \label{fig:L2fig}
\end{figure}

Once the jet injection into the cocoon stops and $Q_{\rm{j}}$ is set to $0$, a new length scale needs to be considered. Using the same approach as \cite{Komissarov_1998} and \cite{Krause_2012}, we derive the expected lobe length for a remnant radio source terminated at length $L_{\rm{on}}$ and time $t_{\rm{on}}$ by considering where the lobe expansion speed is equal to the external sound speed, 
\begin{equation}
L_{2\rm{, rem}} = L_{\rm{on}} \left(\frac{c_{\rm{s, ext}} t_{\rm{on}} }{\alpha L_{\rm{on}}} \right)^{\frac{\alpha}{\alpha-1}}
\end{equation}
\label{eq.lengthscaleremnant}

\noindent where $\alpha$ is the slope of the temporal length evolution for a particular source, $L \propto t^{\alpha}$, and is taken directly from the simulations. Generally, we find a very good agreement between the majority of our simulations and the calculated $L_{2\rm{, rem}}$ value, as shown in Fig. \ref{fig:L2fig}. For high-power simulations in the cluster, the calculated $L_{2\rm{, rem}}$ value agrees remarkably well with the numerical results (see the orange and green lines in the top panel of Fig. \ref{fig:L2fig}).  For those simulations that do not reach the sound speed during the simulated time (i.e. our large switch-off simulations in the group environment; see the red and purple solid lines in the bottom panel of Fig. \ref{fig:L2fig}), the $L_{2, \rm{rem}}$ value is larger than the computational domain. For these simulations, we predict $L_{2, \rm{rem}}$ values of $350$ kpc and $790$ kpc for Q38-v98-a25-G180 and Q36-v01-a25-G180 respectively.\\

Using Eq. \ref{eq.bubblerisevelocity} we again consider the terminal rise speeds of our simulations during the remnant phase. For our simulations in the cluster environment, the forward velocities for 20 kpc and 60 kpc switch-off sources fall below the $300-350$ km/s buoyant velocity in the inner 50 kpc of the cluster profile during the remnant phase. The result is similar for 20 kpc and 60 kpc simulations in the group environment where rise speeds are $\lesssim 250$ km/s. This is consistent with the estimated terminal rise speeds of small bubbles outlined in Section \ref{subsubsec:discussion-buoyancy-active_sources}. Remnant lobes are expected to rise at a velocity below the buoyant velocity once their density approaches that of the surrounding environment, continuing to decelerate as drag becomes significant \citep{Churazov_etal_2000}. This is true of our low-powered sources, where the density of the lobe material becomes increasingly similar to the nearby ambient medium as the lobes mix and entrain surrounding material (we refer the reader to the top two rows of Fig. \ref{fig:remnants}). For high-powered sources, the lobe density is instead a few orders of magnitude lower than the ambient medium at switch-off, however, as these sources remain as a single mass (as opposed to two detached lobes which could independently rise apart) centred at the gravitational potential, buoyancy has a greatly reduced effect on the lobe forward velocities. Instead, the sources drift off the axis of jet propagation.\\

For larger sources which reach 100 kpc or more, Rayleigh-Taylor instabilities begin to pinch the lobe material, enabling dense ambient medium to fall inward towards the gravitational potential. Buoyancy then becomes more important in driving the lobes away from the centre of the grid. \\

The results obtained in this work can be compared to the similar analysis conducted in \cite{Perucho_2014} for their suite of two-dimensional axisymmetric simulations spanning jet powers in the range of $10^{37} - 10^{39}$ W. For their high-powered simulations (e.g. their model J46; a $10^{39}$ W, $v_{\rm{j}}=0.9$c jet), they find the measured advance speed of the cavity to exceed the predicted terminal rise speed by a factor of two; too high to be explained by buoyant rise alone. Conversely for their low-powered sources the measured rise speed sits below the predicted value indicating buoyancy is more likely. In this work, we find strong evidence for the buoyancy-driven expansion of low-powered sources even before the jets have switched off in some cases (our large, low-powered sources). However, for our high-powered sources, we do find terminal rise speeds that match the measured forward advance speeds of the lobe in the case where the environment is more dense (our cluster simulations). 

\subsection{Heating from Remnant Lobes and Shocks?}
\label{subsec:remnant_heaters?}

In this section, we discuss the ability of remnant lobes and shocks to continue to heat the surrounding medium in the remnant phase. This is particularly important as the lobes and shocked regions of our remnant sources tend to sphericity after the jets switch off, offering a mechanism to heat the surrounding medium isotropically. This is feature not seen in active sources due to the inherent directionality of their expansion; \citealp[see also][]{Vernaleo_2006}. \\

Feedback from AGN can act as a thermostat for the host environment, quenching otherwise catastrophic cooling at the centres of clusters \citep{Yang_Reynolds_2016}. The channels through which this heating may occur include cavity heating \citep{Churazov_2001}, the driving of strong shocks \citep{Fabian_2003, Perucho_2014, Shabala_kav_silk_2011}, the dissipation of sound waves, the mixing of thermal gas within the jet-inflated remnant lobes \citep{Hillel_2016}, and the uplifting of low-entropy gas to higher radii by buoyantly rising remnant lobes \citep{Chen_Heinz_Ensslin_2019}. \cite{Shabala_kav_silk_2011} and \cite{Perucho_2014} have shown that the feedback mechanism is dependent on the initial power of the jets. High-powered sources are responsible for the driving of strong shocks, while low-powered sources heat the environment more gently by way of weak shocks and the mixing of the lobes with the ambient medium. Regarding the driver of this mixing, \cite{Bourne_2021} have suggested that cluster weather, rather than the development of Kelvin-Helmholtz and Rayleigh-Taylor instabilities, drives lobe evolution after the lobe-inflation phase. As low-powered jet lobes are more readily disrupted by the cluster weather, they more readily mix with the ICM. On the other hand, \cite{Hillel_2016, Hillel_2017} have argued that it is the development of instabilities that drive mixing between the shocked ambient medium and the lobes.\\

\begin{figure*}
    \centering
    \includegraphics[width=\linewidth]{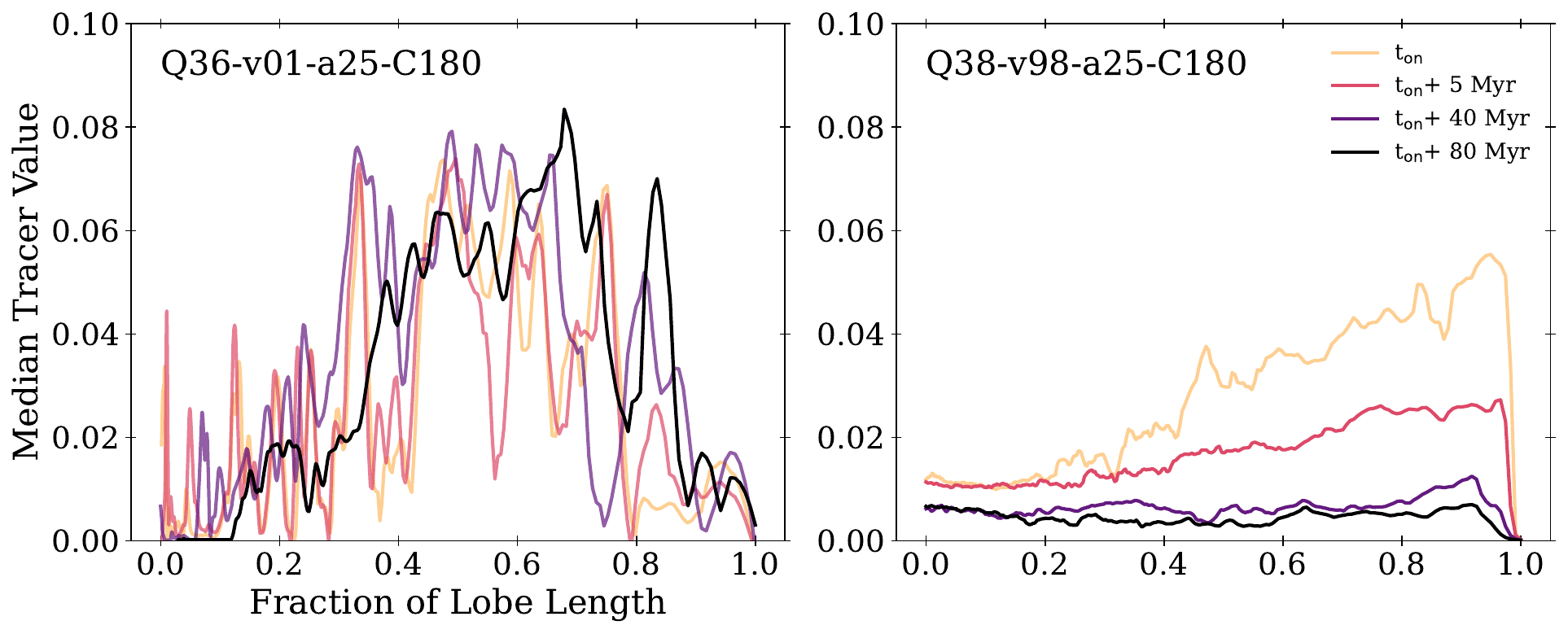}
  \caption{The median tracer value as a function of lobe length along the axis of jet propagation (such that material at the lobe head has a lobe length value of 1) for our largest low-powered simulation in the cluster (Q36-v01-a25-C180, left) and largest high-powered simulation in the cluster (right). We show the tracer distribution at four time steps: at switch-off (yellow), and at 5, 40, and 80 Myr into the remnant phase (pink, purple, black, respectively). The slow changes in tracer values for low-powered sources suggests mixing occurs slowly, while tracer values in the high-powered simulation drop rapidly, indicating fast mixing. }
    \label{fig:tracer_length_plots}
\end{figure*}

The passive tracer advected with the jet fluid is a good indicator of the composition of jet material and ambient material in the lobes of our simulated sources. Considering the change in tracer value as a function of length over time gives an indication of where the mixing is taking place and how rapidly it occurs after the jet switches off. At injection, jet fluid is assigned a tracer value of 1 while the initial ambient medium is assigned 0. Tracer values in between 0 and 1 indicate mixed ambient and jet material. In Fig. \ref{fig:tracer_length_plots} we consider the median tracer value within each row of grid cells along the axis of jet propagation ($\+$ z in our set-up) for two representative simulations; our largest low-powered cluster simulation (left panel of Fig. \ref{fig:tracer_length_plots}) and our largest high-powered cluster simulation (right panel). To isolate the lobe material, we have excluded the recently injected jet material by removing all material with a tracer value greater than 0.2 and above the tracer cutoff of $10^{-4}$. \\

For low-powered sources, the tracer distributions evolve slowly over time and mixing appears to occur preferentially at the base of the lobe. For high-powered sources, tracer values drop rapidly after switch-off, particularly at the head of the lobe. The contact surface is Rayleigh-Taylor unstable due to the deceleration of the denser shocked gas shell with respect to the cocoon \citep{gull_1973}. The instability growth timescale is proportional to the density contrast, which is 2-3 orders of magnitude higher for the more powerful jet. A more detailed study of the post-switch off feedback channels for remnant lobes of high- and low-powered progenitors is deferred to later work.\\

\begin{figure*}
    \centering
    \includegraphics[width=\linewidth]{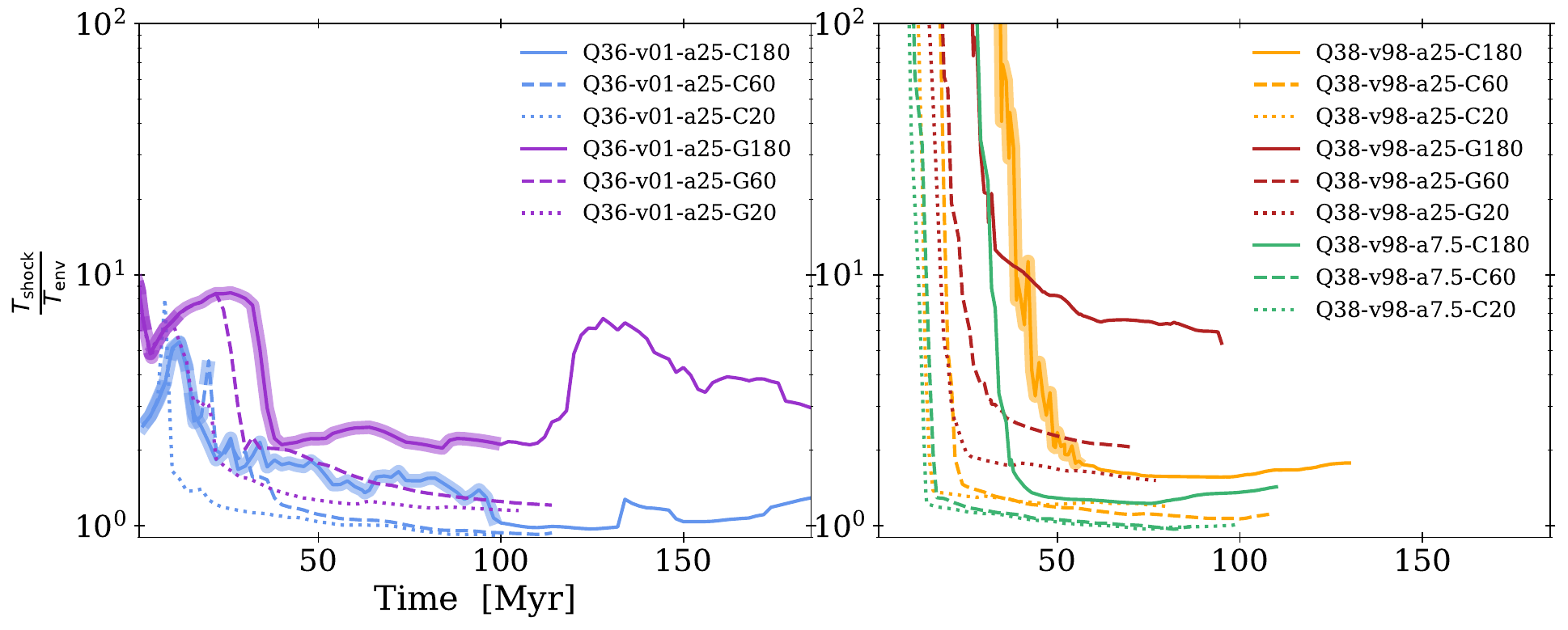}
    \caption{The ratio of bow shock front temperature to environment temperature for all simulations during the active (thick lines) to the remnant phase (thin lines).}
    \label{fig:shock_temperature_ratio}
\end{figure*}

We next consider the heating by the bow shock on the ambient medium. In Fig. \ref{fig:shock_temperature_ratio}, we present the temporal evolution of the ratio of shock temperature to the environment temperature for low- (left panel) and high-powered (right panel) simulations. As with Fig. \ref{fig:shock_properties} in Section \ref{subsec:shocks}, we take the measurement of shock temperature as the maximum value at the leading edge of the shocked region along the propagation axis. For some low-powered simulations (namely, the largest low-power simulations Q36-v01-a25-C180 and Q36-v01-a25-G180), the temperature ratio rises after switch-off, indicating the shock temperature decreases more slowly than the ambient profile.  The temperature ratio for high-powered simulations is consistently higher than low-powered simulations, supporting the idea that the driving of strong shocks is an important heating mechanism for high-powered sources. Further to this, the temperature ratio of high-powered objects plateaus after an initial sharp drop at the start of the remnant phase. Given the significant separation of the shock from the lobe material presented in Section \ref{subsec:shocks}, the shocks of high-powered sources present a way to isotropically heat the surrounding environment at distances up to twice the lobe radii well into the remnant phase. How efficient the feedback is, depends on the evolution of the post-shock gas. For example, relatively flat environments and non-negligible thermal conductivity can both lead to slow cooling of the post-shock gas and hence efficient AGN feedback \citep{Binney_1995, Alexander_2002}. Overall, our results support the idea that both feedback channels are viable. In \cite{Raouf_2017}, both ``shock'' and ``bubble'' modes of RLAGN feedback are considered in a semi-analytic galaxy formation model. They show both are needed to explain the observed galaxy properties, namely the jet-power luminosity relation, stellar mass-radio luminosity relation, and the radio luminosity function as it scales with redshift.

\section{CONCLUSIONS}
\label{sec:conclusions}

We have performed three-dimensional numerical hydrodynamic simulations of relativistic and sub-relativistic AGN jets propagating into realistic environments extracted from cosmological simulations. We evolve these sources from the active phase into the remnant phase by terminating the flow of injected material onto the simulation grid. As an extension to previous works in this area, we consider a parameter space of both low- (Q$_{\rm{j}} = 10^{36}$ W) and high-power (Q$_{\rm{j}} = 10^{38}$ W) jets of varying sizes, in two different environments. The broad dynamics and behaviour of our simulated sources are consistent with earlier work \citep[e.g.][]{Walg_2014, Perucho_2014, English_2019} and can be understood analytically by combining the approaches of \cite{Kaiser_Cotter_2002} and \cite{Turner_2015}. We find the following key results:\\

(i) We confirm several prior results about the dynamic evolution of remnant radio galaxies. We find that some radio remnants from high-power progenitors tend towards more spherical shapes while the leading shock progressively separates from the lobe material. We also confirm the expectation of \cite{Walg_2014} that remnants will experience a delay in their dynamic transition from active to remnant sources. This delay period is set by the time it takes for the trailing jet material to traverse the lobe. For low-powered, slow simulations, this can be as long as 5-10 Myr.\\

(ii) We validate the use of analytic modelling approaches to track the evolution of some remnant radio sources. The lobe and shock evolution of our high-powered, wide jet half-opening angle remnants is well described by analytic approaches if the expansion history is taken into account, allowing for the lobe deceleration and shock separation to be captured correctly. For sources with a narrow jet half-opening angle, the expansion is primarily driven by the momentum of the jet. After switch off, lobe expansion decelerates rapidly, which is not captured in analytic models such as \cite{Kaiser_Cotter_2002}.\\

(iii) We show that analytic approaches can be used effectively to capture the transition from the pressure driven, to the buoyant phase of some simulated remnants. We predict that the overpressure in the lobes of remnants is sufficient to extend the pressure-driven phase by over 100 Myr post switch-off in low density environments.\\

(iv) Remnants of high-power progenitors can supply significant feedback to the ambient gas through bow shock heating, which continues into the remnant phase. We speculate that the feedback supplied by low-powered sources is inefficient, with mixing between the shock and lobe occurring very gradually.\\

Detailed predictions of observable synchrotron emission are required to interpret observations of remnant candidates, and complement the dynamical study presented here. We defer this to the companion work in \cite{Stewart_2025}.

\section{Acknowledgements}

We thank the anonymous referee for their helpful comments which added value to the manuscript. GS thanks the Australian Government for an Australian Government Research Training Program RTP scholarship and the CSIRO for an ATNF studentship. SS, CP and PYJ acknowledge the Australian Research Council grant DP240102970. This research was carried out using high-performance computing infrastructure provided by the Tasmanian Partnership for Advanced Computing (TPAC). We acknowledge Geli Kourakis and John Miezitis who provided valuable assistance throughout this project. We acknowledge the support of the developers providing the Python packages that made this work possible: JupyterLab \citep{Kluyver_Jupyter_Lab}, Matplotlib \citep{Hunter_matplotlib}, NumPy \citep{Harris_numpy}, SciPy \citep{Virtanen_Scipy}, and Astropy \citep{Robitaille_2013_astropy}.



\bibliography{references}

\begin{thebibliography}{}
\expandafter\ifx\csname natexlab\endcsname\relax\def\natexlab#1{#1}\fi

\bibitem[{Alexander(2002)}]{Alexander_2002}
Alexander, P. 2002, Monthly Notices of the Royal Astronomical Society, 335, 610–620

\bibitem[{Alexander(2006)}]{Alexander_2006}
---. 2006, Monthly Notices of the Royal Astronomical Society, 368, 1404–1410

\bibitem[{An \& Baan(2012)}]{An_Baan_2012}
An, T., \& Baan, W.~A. 2012, The Astrophysical Journal, 760, 77

\bibitem[{Barišić {et~al.}(2017)Barišić, van~der Wel, Bezanson, Pacifici, Noeske, Muñoz-Mateos, Franx, Smolčić, Bell, Brammer, Calhau, Chauké, van Dokkum, van Houdt, Gallazzi, Labbé, Maseda, Muzzin, Sobral, Straatman, \& Wu}]{Bari_i__2017}
Barišić, I., van~der Wel, A., Bezanson, R., {et~al.} 2017, The Astrophysical Journal, 847, 72

\bibitem[{Best {et~al.}(2005)Best, Kauffmann, Heckman, Brinchmann, Charlot, Ivezić, \& White}]{Best_2005}
Best, P.~N., Kauffmann, G., Heckman, T.~M., {et~al.} 2005, Monthly Notices of the Royal Astronomical Society, 362, 25–40

\bibitem[{Best {et~al.}(2007)Best, Von Der~Linden, Kauffmann, Heckman, \& Kaiser}]{Best_2007}
Best, P.~N., Von Der~Linden, A., Kauffmann, G., Heckman, T.~M., \& Kaiser, C.~R. 2007, Monthly Notices of the Royal Astronomical Society, 379, 894–908

\bibitem[{Binney \& Tabor(1995)}]{Binney_1995}
Binney, J., \& Tabor, G. 1995, Monthly Notices of the Royal Astronomical Society, 276, 663–678

\bibitem[{Bird {et~al.}(2022)Bird, Ni, Di~Matteo, Croft, Feng, \& Chen}]{Bird_2022}
Bird, S., Ni, Y., Di~Matteo, T., {et~al.} 2022, Monthly Notices of the Royal Astronomical Society, 512, 3703–3716

\bibitem[{Boehringer {et~al.}(1993)Boehringer, Voges, Fabian, Edge, \& Neumann}]{Boehringer_1993}
Boehringer, H., Voges, W., Fabian, A.~C., Edge, A.~C., \& Neumann, D.~M. 1993, Monthly Notices of the Royal Astronomical Society, 264, L25–L28

\bibitem[{Bourne \& Sijacki(2021)}]{Bourne_2021}
Bourne, M.~A., \& Sijacki, D. 2021, Monthly Notices of the Royal Astronomical Society, 506, 488–513

\bibitem[{Brienza {et~al.}(2016)Brienza, Godfrey, Morganti, Vilchez, Maddox, Murgia, Orru, Shulevski, Best, Brüggen, Harwood, Jamrozy, Jarvis, Mahony, McKean, \& Röttgering}]{Brienza_2016}
Brienza, M., Godfrey, L., Morganti, R., {et~al.} 2016, Astronomy \&; Astrophysics, 585, A29

\bibitem[{Brienza {et~al.}(2017)Brienza, Godfrey, Morganti, Prandoni, Harwood, Mahony, Hardcastle, Murgia, Röttgering, Shimwell, \& Shulevski}]{Brienza_2017}
---. 2017, Astronomy \&; Astrophysics, 606, A98

\bibitem[{Brocksopp {et~al.}(2011)Brocksopp, Kaiser, Schoenmakers, \& de~Bruyn}]{Brocksopp_2011}
Brocksopp, C., Kaiser, C.~R., Schoenmakers, A.~P., \& de~Bruyn, A.~G. 2011, Monthly Notices of the Royal Astronomical Society, 410, 484–498

\bibitem[{Chen {et~al.}(2019)Chen, Heinz, \& Enßlin}]{Chen_Heinz_Ensslin_2019}
Chen, Y.-H., Heinz, S., \& Enßlin, T.~A. 2019, Monthly Notices of the Royal Astronomical Society, 489, 1939–1949

\bibitem[{Churazov {et~al.}(2001)Churazov, Bruggen, Kaiser, Bohringer, \& Forman}]{Churazov_2001}
Churazov, E., Bruggen, M., Kaiser, C.~R., Bohringer, H., \& Forman, W. 2001, The Astrophysical Journal, 554, 261–273

\bibitem[{Churazov {et~al.}(2000)Churazov, W.Forman, C., \& Bohringer}]{Churazov_etal_2000}
Churazov, E., W.Forman, W., C., J., \& Bohringer, H. 2000, Astronomy and Astrophysics, 356, 788

\bibitem[{Croft {et~al.}(2006)Croft, van Breugel, de~Vries, Dopita, Martin, Morganti, Neff, Oosterloo, Schiminovich, Stanford, \& van Gorkom}]{Croft_2006}
Croft, S., van Breugel, W., de~Vries, W., {et~al.} 2006, The Astrophysical Journal, 647, 1040

\bibitem[{Cui {et~al.}(2018)Cui, Knebe, Yepes, Pearce, Power, Dave, Arth, Borgani, Dolag, Elahi, Mostoghiu, Murante, Rasia, Stoppacher, Vega-Ferrero, Wang, Yang, Benson, Cora, Croton, Sinha, Stevens, Vega-Martínez, Arthur, Baldi, Cañas, Cialone, Cunnama, De~Petris, Durando, Ettori, Gottlöber, Nuza, Old, Pilipenko, Sorce, \& Welker}]{Cui_2018}
Cui, W., Knebe, A., Yepes, G., {et~al.} 2018, Monthly Notices of the Royal Astronomical Society, 480, 2898–2915

\bibitem[{Davé {et~al.}(2019)Davé, Anglés-Alcázar, Narayanan, Li, Rafieferantsoa, \& Appleby}]{Dav__2019}
Davé, R., Anglés-Alcázar, D., Narayanan, D., {et~al.} 2019, Monthly Notices of the Royal Astronomical Society, 486, 2827–2849

\bibitem[{Dugan {et~al.}(2017)Dugan, Gaibler, \& Silk}]{Dugan_2017}
Dugan, Z., Gaibler, V., \& Silk, J. 2017, The Astrophysical Journal, 844, 37

\bibitem[{Dutta {et~al.}(2023)Dutta, Singh, Chandra, Wadadekar, Kayal, \& Heywood}]{Dutta_2023}
Dutta, S., Singh, V., Chandra, C. H.~I., {et~al.} 2023, The Astrophysical Journal, 944, 176

\bibitem[{English {et~al.}(2019)English, Hardcastle, \& Krause}]{English_2019}
English, W., Hardcastle, M.~J., \& Krause, M. G.~H. 2019, Monthly Notices of the Royal Astronomical Society, 490, 5807

\bibitem[{Enßlin \& Heinz(2002)}]{Ensslin_2002}
Enßlin, T.~A., \& Heinz, S. 2002, Astronomy \&; Astrophysics, 384, L27–L30

\bibitem[{Fabian(2012)}]{Fabian_2012}
Fabian, A. 2012, Annual Review of Astronomy and Astrophysics, 50, 455–489

\bibitem[{Fabian {et~al.}(2003)Fabian, Sanders, Allen, Crawford, Iwasawa, Johnstone, Schmidt, \& Taylor}]{Fabian_2003}
Fabian, A.~C., Sanders, J.~S., Allen, S.~W., {et~al.} 2003, Monthly Notices of the Royal Astronomical Society, 344, L43–L47

\bibitem[{{Fanti} {et~al.}(1995){Fanti}, {Fanti}, {Dallacasa}, {Schilizzi}, {Spencer}, \& {Stanghellini}}]{Fanti_1995}
{Fanti}, C., {Fanti}, R., {Dallacasa}, D., {et~al.} 1995, Astronomy \& Astophysics, 302, 317

\bibitem[{Forman {et~al.}(2005)Forman, Nulsen, Heinz, Owen, Eilek, Vikhlinin, Markevitch, Kraft, Churazov, \& Jones}]{Forman_2005}
Forman, W., Nulsen, P., Heinz, S., {et~al.} 2005, The Astrophysical Journal, 635, 894–906

\bibitem[{Giovannini {et~al.}(2001)Giovannini, Cotton, Feretti, Lara, \& Venturi}]{Giovannini_2001}
Giovannini, G., Cotton, W.~D., Feretti, L., Lara, L., \& Venturi, T. 2001, The Astrophysical Journal, 552, 508–526

\bibitem[{Giroletti \& Polatidis(2009)}]{Giroletti_2009}
Giroletti, M., \& Polatidis, A. 2009, Astronomische Nachrichten, 330, 193–198

\bibitem[{Gull \& Northover(1973)}]{gull_1973}
Gull, S.~F., \& Northover, K. J.~E. 1973, Nature, 244, 80–83

\bibitem[{Hardcastle(2018)}]{Hardcastle_2018}
Hardcastle, M.~J. 2018, Monthly Notices of the Royal Astronomical Society, 475, 2768

\bibitem[{Hardcastle \& Krause(2013)}]{Hardcastle_Krause_2013}
Hardcastle, M.~J., \& Krause, M. G.~H. 2013, Monthly Notices of the Royal Astronomical Society, 430, 174

\bibitem[{Hardcastle {et~al.}(2019)Hardcastle, Williams, Best, Croston, Duncan, Röttgering, Sabater, Shimwell, Tasse, Callingham, Cochrane, de~Gasperin, Gürkan, Jarvis, Mahatma, Miley, Mingo, Mooney, Morabito, O’Sullivan, Prandoni, Shulevski, \& Smith}]{Hardcastle_2019}
Hardcastle, M.~J., Williams, W.~L., Best, P.~N., {et~al.} 2019, Astronomy \&; Astrophysics, 622, A12

\bibitem[{{Harris} {et~al.}(2020){Harris}, {Millman}, {van der Walt}, {Gommers}, {Virtanen}, {Cournapeau}, {Wieser}, {Taylor}, {Berg}, {Smith}, {Kern}, {Picus}, {Hoyer}, {van Kerkwijk}, {Brett}, {Haldane}, {del R{\'\i}o}, {Wiebe}, {Peterson}, {G{\'e}rard-Marchant}, {Sheppard}, {Reddy}, {Weckesser}, {Abbasi}, {Gohlke}, \& {Oliphant}}]{Harris_numpy}
{Harris}, C.~R., {Millman}, K.~J., {van der Walt}, S.~J., {et~al.} 2020, \nat, 585, 357

\bibitem[{Hillel \& Soker(2016)}]{Hillel_2016}
Hillel, S., \& Soker, N. 2016, Monthly Notices of the Royal Astronomical Society, 455, 2139–2148

\bibitem[{Hillel \& Soker(2017)}]{Hillel_2017}
---. 2017, The Astrophysical Journal, 845, 91

\bibitem[{{Hunter}(2007)}]{Hunter_matplotlib}
{Hunter}, J.~D. 2007, Computing in Science and Engineering, 9, 90

\bibitem[{Jurlin {et~al.}(2021)Jurlin, Brienza, Morganti, Wadadekar, Ishwara-Chandra, Maddox, \& Mahatma}]{Jurlin_2021}
Jurlin, N., Brienza, M., Morganti, R., {et~al.} 2021, Astronomy \&; Astrophysics, 653, A110

\bibitem[{Jurlin {et~al.}(2020)Jurlin, Morganti, Brienza, Mandal, Maddox, Duncan, Shabala, Hardcastle, Prandoni, Röttgering, Mahatma, Best, Mingo, Sabater, Shimwell, \& Tasse}]{Jurlin_2020}
Jurlin, N., Morganti, R., Brienza, M., {et~al.} 2020, Astronomy \&; Astrophysics, 638, A34

\bibitem[{Kaiser \& Alexander(1997)}]{Kaiser_Alexander_1997}
Kaiser, C.~R., \& Alexander, P. 1997, Monthly Notices of the Royal Astronomical Society, 286, 215

\bibitem[{Kaiser \& Cotter(2002)}]{Kaiser_Cotter_2002}
Kaiser, C.~R., \& Cotter, G. 2002, Monthly Notices of the Royal Astronomical Society, 336, 649

\bibitem[{Kawata \& Gibson(2005)}]{Kawata_2005}
Kawata, D., \& Gibson, B.~K. 2005, Monthly Notices of the Royal Astronomical Society: Letters, 358, L16–L20

\bibitem[{{Kluyver} {et~al.}(2016){Kluyver}, {Ragan-Kelley}, {P{\'e}rez}, {Granger}, {Bussonnier}, {Frederic}, {Kelley}, {Hamrick}, {Grout}, {Corlay}, {Ivanov}, {Avila}, {Abdalla}, {Willing}, \& {Jupyter Development Team}}]{Kluyver_Jupyter_Lab}
{Kluyver}, T., {Ragan-Kelley}, B., {P{\'e}rez}, F., {et~al.} 2016, in IOS Press, 87--90

\bibitem[{Komissarov \& Falle(1998)}]{Komissarov_1998}
Komissarov, S.~S., \& Falle, S. A. E.~G. 1998, Monthly Notices of the Royal Astronomical Society, 297, 1087–1108

\bibitem[{Krause(2005)}]{Krause_2005}
Krause, M. 2005, Astronomy \&; Astrophysics, 431, 45–64

\bibitem[{Krause {et~al.}(2012)Krause, Alexander, Riley, \& Hopton}]{Krause_2012}
Krause, M., Alexander, P., Riley, J., \& Hopton, D. 2012, Monthly Notices of the Royal Astronomical Society, 427, 3196–3208

\bibitem[{Laing \& Bridle(2014)}]{Laing_2014}
Laing, R.~A., \& Bridle, A.~H. 2014, Monthly Notices of the Royal Astronomical Society, 437, 3405–3441

\bibitem[{Li {et~al.}(2021)Li, Ni, Croft, Di~Matteo, Bird, \& Feng}]{Li_2021}
Li, Y., Ni, Y., Croft, R. A.~C., {et~al.} 2021, Proceedings of the National Academy of Sciences, 118

\bibitem[{Mahatma(2023)}]{Mahatma_2023}
Mahatma, V.~H. 2023, Galaxies, 11, 74

\bibitem[{Mahatma {et~al.}(2018)Mahatma, Hardcastle, Williams, Brienza, Brüggen, Croston, Gurkan, Harwood, Kunert-Bajraszewska, Morganti, Röttgering, Shimwell, \& Tasse}]{Mahatma_2018}
Mahatma, V.~H., Hardcastle, M.~J., Williams, W.~L., {et~al.} 2018, Monthly Notices of the Royal Astronomical Society, 475, 4557–4578

\bibitem[{Mahony {et~al.}(2015)Mahony, Oonk, Morganti, Tadhunter, Bessiere, Short, Emonts, \& Oosterloo}]{Mahony_2015}
Mahony, E.~K., Oonk, J. B.~R., Morganti, R., {et~al.} 2015, Monthly Notices of the Royal Astronomical Society, 455, 2453

\bibitem[{Mandal {et~al.}(2021)Mandal, Mukherjee, Federrath, Nesvadba, Bicknell, Wagner, \& Meenakshi}]{Mandal_2021}
Mandal, A., Mukherjee, D., Federrath, C., {et~al.} 2021, Monthly Notices of the Royal Astronomical Society, 508, 4738–4757

\bibitem[{Massaglia {et~al.}(2016)Massaglia, Bodo, Rossi, Capetti, \& Mignone}]{Massaglia_2016}
Massaglia, S., Bodo, G., Rossi, P., Capetti, S., \& Mignone, A. 2016, Astronomy \& Astrophysics, 596, A12

\bibitem[{Mathews(1971)}]{Mathews_1971}
Mathews, W.~G. 1971, The Astrophysical Journal, 165, 147

\bibitem[{Mendygral {et~al.}(2012)Mendygral, Jones, \& Dolag}]{Mendygral_2012}
Mendygral, P.~J., Jones, T.~W., \& Dolag, K. 2012, The Astrophysical Journal, 750, 166

\bibitem[{Mignone \& McKinney(2007)}]{Mignone_McKinney2007}
Mignone, A., \& McKinney, J.~C. 2007, Monthly Notices of the Royal Astronomical Society, 378, 1118–1130

\bibitem[{Mittal {et~al.}(2009)Mittal, Hudson, Reiprich, \& Clarke}]{Mittal_2009}
Mittal, R., Hudson, D.~S., Reiprich, T.~H., \& Clarke, T. 2009, Astronomy \&; Astrophysics, 501, 835–850

\bibitem[{{Monaghan} \& {Lattanzio}(1985)}]{Monaghan_1985}
{Monaghan}, J.~J., \& {Lattanzio}, J.~C. 1985, Astronomy and Astrophysics, 149, 135

\bibitem[{Morganti {et~al.}(2013)Morganti, Fogasy, Paragi, Oosterloo, \& Orienti}]{Morganti_2013}
Morganti, R., Fogasy, J., Paragi, Z., Oosterloo, T., \& Orienti, M. 2013, Science, 341, 1082

\bibitem[{Morganti {et~al.}(2005)Morganti, Tadhunter, \& Oosterloo}]{Morganti_2005}
Morganti, R., Tadhunter, C.~N., \& Oosterloo, T.~A. 2005, Astronomy \& Astrophysics, 444, L9

\bibitem[{Murgia(2003)}]{Murgia_2003}
Murgia, M. 2003, Publications of the Astronomical Society of Australia, 20, 19–24

\bibitem[{Murgia {et~al.}(2011)Murgia, Parma, Mack, de~Ruiter, Fanti, Govoni, Tarchi, Giacintucci, \& Markevitch}]{Murgia_2011}
Murgia, M., Parma, P., Mack, K.-H., {et~al.} 2011, Astronomy \&; Astrophysics, 526, A148

\bibitem[{Nandi {et~al.}(2019)Nandi, Saikia, Roy, Dabhade, Wadadekar, Larsson, Baes, Chandola, \& Singh}]{Nandi_2019}
Nandi, S., Saikia, D.~J., Roy, R., {et~al.} 2019, Monthly Notices of the Royal Astronomical Society, 486, 5158–5170

\bibitem[{Nesvadba {et~al.}(2020)Nesvadba, Bicknell, Mukherjee, \& Wagner}]{Nesvadba_2020}
Nesvadba, N. P.~H., Bicknell, G.~V., Mukherjee, D., \& Wagner, A.~Y. 2020, Astronomy \&; Astrophysics, 639, L13

\bibitem[{Nesvadba {et~al.}(2011)Nesvadba, Boulanger, Lehnert, Guillard, \& Salome}]{Nesvadba_2011}
Nesvadba, N. P.~H., Boulanger, F., Lehnert, M.~D., Guillard, P., \& Salome, P. 2011, Astronomy \&; Astrophysics, 536, L5

\bibitem[{Nesvadba {et~al.}(2010)Nesvadba, Boulanger, Salomé, Guillard, Lehnert, Ogle, Appleton, Falgarone, \& Pineau~des Forets}]{Nesvadba_2010}
Nesvadba, N. P.~H., Boulanger, F., Salomé, P., {et~al.} 2010, Astronomy and Astrophysics, 521, A65

\bibitem[{Novak {et~al.}(2011)Novak, Ostriker, \& Ciotti}]{Novak_2011}
Novak, G.~S., Ostriker, J.~P., \& Ciotti, L. 2011, The Astrophysical Journal, 737, 26

\bibitem[{Orrù {et~al.}(2010)Orrù, Murgia, Feretti, Govoni, Giovannini, Lane, Kassim, \& Paladino}]{Orru_2010}
Orrù, E., Murgia, M., Feretti, L., {et~al.} 2010, Astronomy and Astrophysics, 515, A50

\bibitem[{Orrù {et~al.}(2015)Orrù, van Velzen, Pizzo, Yatawatta, Paladino, Iacobelli, Murgia, Falcke, Morganti, de~Bruyn, Ferrari, Anderson, Bonafede, Mulcahy, Asgekar, Avruch, Beck, Bell, van Bemmel, Bentum, Bernardi, Best, Breitling, Broderick, Brüggen, Butcher, Ciardi, Conway, Corstanje, de~Geus, Deller, Duscha, Eislöffel, Engels, Frieswijk, Garrett, Grießmeier, Gunst, Hamaker, Heald, Hoeft, van~der Horst, Intema, Juette, Kohler, Kondratiev, Kuniyoshi, Kuper, Loose, Maat, Mann, Markoff, McFadden, McKay-Bukowski, Miley, Moldon, Molenaar, Munk, Nelles, Paas, Pandey-Pommier, Pandey, Pietka, Polatidis, Reich, Röttgering, Rowlinson, Scaife, Schoenmakers, Schwarz, Serylak, Shulevski, Smirnov, Steinmetz, Stewart, Swinbank, Tagger, Tasse, Thoudam, Toribio, Vermeulen, Vocks, van Weeren, Wijers, Wise, \& Wucknitz}]{Orru_2015}
Orrù, E., van Velzen, S., Pizzo, R.~F., {et~al.} 2015, Astronomy \&; Astrophysics, 584, A112

\bibitem[{O’Dea(1998)}]{O_Dea_1998}
O’Dea, C.~P. 1998, Publications of the Astronomical Society of the Pacific, 110, 493–532

\bibitem[{Parma {et~al.}(2007)Parma, Murgia, de~Ruiter, Fanti, Mack, \& Govoni}]{Parma_2007}
Parma, P., Murgia, M., de~Ruiter, H.~R., {et~al.} 2007, Astronomy \&; Astrophysics, 470, 875–888

\bibitem[{Perucho {et~al.}(2014)Perucho, Martí, Quilis, \& Ricciardelli}]{Perucho_2014}
Perucho, M., Martí, J.-M., Quilis, V., \& Ricciardelli, E. 2014, Monthly Notices of the Royal Astronomical Society, 445, 1462–1481

\bibitem[{Perucho {et~al.}(2011)Perucho, Quilis, \& Mart{\'{\i}}}]{Perucho_2011}
Perucho, M., Quilis, V., \& Mart{\'{\i}}, J.-M. 2011, The Astrophysical Journal, 743, 42

\bibitem[{{Planck Collaboration} {et~al.}(2016){Planck Collaboration}, {Ade}, {Aghanim}, {Arnaud}, {Ashdown}, {Aumont}, {Baccigalupi}, {Banday}, {Barreiro}, {Bartlett}, {Bartolo}, {Battaner}, {Battye}, {Benabed}, {Beno{\^\i}t}, {Benoit-L{\'e}vy}, {Bernard}, {Bersanelli}, {Bielewicz}, {Bock}, {Bonaldi}, {Bonavera}, {Bond}, {Borrill}, {Bouchet}, {Boulanger}, {Bucher}, {Burigana}, {Butler}, {Calabrese}, {Cardoso}, {Catalano}, {Challinor}, {Chamballu}, {Chary}, {Chiang}, {Chluba}, {Christensen}, {Church}, {Clements}, {Colombi}, {Colombo}, {Combet}, {Coulais}, {Crill}, {Curto}, {Cuttaia}, {Danese}, {Davies}, {Davis}, {de Bernardis}, {de Rosa}, {de Zotti}, {Delabrouille}, {D{\'e}sert}, {Di Valentino}, {Dickinson}, {Diego}, {Dolag}, {Dole}, {Donzelli}, {Dor{\'e}}, {Douspis}, {Ducout}, {Dunkley}, {Dupac}, {Efstathiou}, {Elsner}, {En{\ss}lin}, {Eriksen}, {Farhang}, {Fergusson}, {Finelli}, {Forni}, {Frailis}, {Fraisse}, {Franceschi}, {Frejsel}, {Galeotta}, {Galli}, {Ganga}, {Gauthier}, {Gerbino}, {Ghosh}, {Giard},
  {Giraud-H{\'e}raud}, {Giusarma}, {Gjerl{\o}w}, {Gonz{\'a}lez-Nuevo}, {G{\'o}rski}, {Gratton}, {Gregorio}, {Gruppuso}, {Gudmundsson}, {Hamann}, {Hansen}, {Hanson}, {Harrison}, {Helou}, {Henrot-Versill{\'e}}, {Hern{\'a}ndez-Monteagudo}, {Herranz}, {Hildebrandt}, {Hivon}, {Hobson}, {Holmes}, {Hornstrup}, {Hovest}, {Huang}, {Huffenberger}, {Hurier}, {Jaffe}, {Jaffe}, {Jones}, {Juvela}, {Keih{\"a}nen}, {Keskitalo}, {Kisner}, {Kneissl}, {Knoche}, {Knox}, {Kunz}, {Kurki-Suonio}, {Lagache}, {L{\"a}hteenm{\"a}ki}, {Lamarre}, {Lasenby}, {Lattanzi}, {Lawrence}, {Leahy}, {Leonardi}, {Lesgourgues}, {Levrier}, {Lewis}, {Liguori}, {Lilje}, {Linden-V{\o}rnle}, {L{\'o}pez-Caniego}, {Lubin}, {Mac{\'\i}as-P{\'e}rez}, {Maggio}, {Maino}, {Mandolesi}, {Mangilli}, {Marchini}, {Maris}, {Martin}, {Martinelli}, {Mart{\'\i}nez-Gonz{\'a}lez}, {Masi}, {Matarrese}, {McGehee}, {Meinhold}, {Melchiorri}, {Melin}, {Mendes}, {Mennella}, {Migliaccio}, {Millea}, {Mitra}, {Miville-Desch{\^e}nes}, {Moneti}, {Montier}, {Morgante}, {Mortlock},
  {Moss}, {Munshi}, {Murphy}, {Naselsky}, {Nati}, {Natoli}, {Netterfield}, {N{\o}rgaard-Nielsen}, {Noviello}, {Novikov}, {Novikov}, {Oxborrow}, {Paci}, {Pagano}, {Pajot}, {Paladini}, {Paoletti}, {Partridge}, {Pasian}, {Patanchon}, {Pearson}, {Perdereau}, {Perotto}, {Perrotta}, {Pettorino}, {Piacentini}, {Piat}, {Pierpaoli}, {Pietrobon}, {Plaszczynski}, {Pointecouteau}, {Polenta}, {Popa}, {Pratt}, {Pr{\'e}zeau}, {Prunet}, {Puget}, {Rachen}, {Reach}, {Rebolo}, {Reinecke}, {Remazeilles}, {Renault}, {Renzi}, {Ristorcelli}, {Rocha}, {Rosset}, {Rossetti}, {Roudier}, {Rouill{\'e} d'Orfeuil}, {Rowan-Robinson}, {Rubi{\~n}o-Mart{\'\i}n}, {Rusholme}, {Said}, {Salvatelli}, {Salvati}, {Sandri}, {Santos}, {Savelainen}, {Savini}, {Scott}, {Seiffert}, {Serra}, {Shellard}, {Spencer}, {Spinelli}, {Stolyarov}, {Stompor}, {Sudiwala}, {Sunyaev}, {Sutton}, {Suur-Uski}, {Sygnet}, {Tauber}, {Terenzi}, {Toffolatti}, {Tomasi}, {Tristram}, {Trombetti}, {Tucci}, {Tuovinen}, {T{\"u}rler}, {Umana}, {Valenziano}, {Valiviita}, {Van Tent},
  {Vielva}, {Villa}, {Wade}, {Wandelt}, {Wehus}, {White}, {White}, {Wilkinson}, {Yvon}, {Zacchei}, \& {Zonca}}]{Plank_Collab_2016}
{Planck Collaboration}, {Ade}, P.~A.~R., {Aghanim}, N., {et~al.} 2016, \aap, 594, A13

\bibitem[{Polatidis \& Conway(2003)}]{Polatidis_2003}
Polatidis, A.~G., \& Conway, J.~E. 2003, Publications of the Astronomical Society of Australia, 20, 69–74

\bibitem[{Pope {et~al.}(2012)Pope, Mendel, \& Shabala}]{Pope_2011}
Pope, E. C.~D., Mendel, J.~T., \& Shabala, S.~S. 2012, Monthly Notices of the Royal Astronomical Society, 419, 50–56

\bibitem[{Quici {et~al.}(2025)Quici, Turner, Seymour, \& Hurley-Walker}]{Quici_2025}
Quici, B., Turner, R.~J., Seymour, N., \& Hurley-Walker, N. 2025, Monthly Notices of the Royal Astronomical Society, 537, 343–363

\bibitem[{Quici {et~al.}(2021)Quici, Hurley-Walker, Seymour, Turner, Shabala, Huynh, Andernach, Kapińska, Collier, Johnston-Hollitt, White, Prandoni, Galvin, Franzen, Ishwara-Chandra, Bellstedt, Tingay, Gaensler, O’Brien, Rogers, Chow, Driver, \& Robotham}]{Quici_2021}
Quici, B., Hurley-Walker, N., Seymour, N., {et~al.} 2021, Publications of the Astronomical Society of Australia, 38

\bibitem[{Raouf {et~al.}(2017)Raouf, Shabala, Croton, Khosroshahi, \& Bernyk}]{Raouf_2017}
Raouf, M., Shabala, S.~S., Croton, D.~J., Khosroshahi, H.~G., \& Bernyk, M. 2017, Monthly Notices of the Royal Astronomical Society, 471, 658–670

\bibitem[{Reynolds {et~al.}(2005)Reynolds, McKernan, Fabian, Stone, \& Vernaleo}]{Reynolds_2005}
Reynolds, C.~S., McKernan, B., Fabian, A.~C., Stone, J.~M., \& Vernaleo, J.~C. 2005, Monthly Notices of the Royal Astronomical Society, 357, 242–250

\bibitem[{Robitaille {et~al.}(2013)Robitaille, Tollerud, Greenfield, Droettboom, Bray, \& Aldcroft}]{Robitaille_2013_astropy}
Robitaille, T., Tollerud, E.~J., Greenfield, P., {et~al.} 2013, Astronomy \& Astrophysics

\bibitem[{Rodman {et~al.}(2018)Rodman, Turner, Shabala, Banfield, Wong, Andernach, Garon, Kapińska, Norris, \& Rudnick}]{Rodman_2018}
Rodman, P.~E., Turner, R.~J., Shabala, S.~S., {et~al.} 2018, Monthly Notices of the Royal Astronomical Society, 482, 5625–5641

\bibitem[{Sabater {et~al.}(2019)Sabater, Best, Hardcastle, Shimwell, Tasse, Williams, Brüggen, Cochrane, Croston, de~Gasperin, Duncan, Gürkan, Mechev, Morabito, Prandoni, Röttgering, Smith, Harwood, Mingo, Mooney, \& Saxena}]{Sabater_2019}
Sabater, J., Best, P.~N., Hardcastle, M.~J., {et~al.} 2019, Astronomy \&; Astrophysics, 622, A17

\bibitem[{Santoro {et~al.}(2020)Santoro, Tadhunter, Baron, Morganti, \& Holt}]{Santoro_2020}
Santoro, F., Tadhunter, C., Baron, D., Morganti, R., \& Holt, J. 2020, Astronomy \& Astrophysics, 644, A54

\bibitem[{Schaye {et~al.}(2014)Schaye, Crain, Bower, Furlong, Schaller, Theuns, Dalla~Vecchia, Frenk, McCarthy, Helly, Jenkins, Rosas-Guevara, White, Baes, Booth, Camps, Navarro, Qu, Rahmati, Sawala, Thomas, \& Trayford}]{Schaye_2014}
Schaye, J., Crain, R.~A., Bower, R.~G., {et~al.} 2014, Monthly Notices of the Royal Astronomical Society, 446, 521–554

\bibitem[{Schoenmakers {et~al.}(2000)Schoenmakers, de~Bruyn, Rottgering, van~der Laan, \& Kaiser}]{Schoenmakers_et_al_2000}
Schoenmakers, A.~P., de~Bruyn, A.~G., Rottgering, H. J.~A., van~der Laan, H., \& Kaiser, C.~R. 2000, Monthly Notices of the Royal Astronomical Society, 315, 371

\bibitem[{Seabold \& Perktold(2010)}]{statsmodels}
Seabold, S., \& Perktold, J. 2010, in 9th Python in Science Conference

\bibitem[{Shabala \& Alexander(2009{\natexlab{a}})}]{Shabala_Alexander_feedback_2009}
Shabala, S., \& Alexander, P. 2009{\natexlab{a}}, The Astrophysical Journal, 699, 525–538

\bibitem[{Shabala \& Alexander(2009{\natexlab{b}})}]{Shabala_sound_waves_2009}
Shabala, S.~S., \& Alexander, P. 2009{\natexlab{b}}, Monthly Notices of the Royal Astronomical Society, 392, 1413–1420

\bibitem[{Shabala {et~al.}(2020)Shabala, Jurlin, Morganti, Brienza, Hardcastle, Godfrey, Krause, \& Turner}]{Shabala_2020}
Shabala, S.~S., Jurlin, N., Morganti, R., {et~al.} 2020, Monthly Notices of the Royal Astronomical Society, 496, 1706–1717

\bibitem[{Shabala {et~al.}(2011)Shabala, Kaviraj, \& Silk}]{Shabala_kav_silk_2011}
Shabala, S.~S., Kaviraj, S., \& Silk, J. 2011, Monthly Notices of the Royal Astronomical Society, 413, 2815–2826

\bibitem[{Shulevski {et~al.}(2015)Shulevski, Morganti, Barthel, Murgia, van Weeren, White, Brüggen, Kunert-Bajraszewska, Jamrozy, Best, Röttgering, Chyzy, de~Gasperin, Bîrzan, Brunetti, Brienza, Rafferty, Anderson, Beck, Deller, Zarka, Schwarz, Mahony, Orrú, Bell, Bentum, Bernardi, Bonafede, Breitling, Broderick, Butcher, Carbone, Ciardi, de~Geus, Duscha, Eislöffel, Engels, Falcke, Fallows, Fender, Ferrari, Frieswijk, Garrett, Grießmeier, Gunst, Heald, Hoeft, Hörandel, Horneffer, van~der Horst, Intema, Juette, Karastergiou, Kondratiev, Kramer, Kuniyoshi, Kuper, Maat, Mann, McFadden, McKay-Bukowski, McKean, Meulman, Mulcahy, Munk, Norden, Paas, Pandey-Pommier, Pizzo, Polatidis, Reich, Rowlinson, Scaife, Serylak, Sluman, Smirnov, Steinmetz, Swinbank, Tagger, Tang, Tasse, Thoudam, Toribio, Vermeulen, Vocks, Wijers, Wise, \& Wucknitz}]{Shulevski_2015}
Shulevski, A., Morganti, R., Barthel, P.~D., {et~al.} 2015, Astronomy \& Astrophysics, 579, A27

\bibitem[{Sijacki {et~al.}(2007)Sijacki, Springel, Di~Matteo, \& Hernquist}]{Sijacki_2007}
Sijacki, D., Springel, V., Di~Matteo, T., \& Hernquist, L. 2007, Monthly Notices of the Royal Astronomical Society, 380, 877–900

\bibitem[{Stewart {et~al.}(in prep.)Stewart, Shabala, Yates-Jones, Turner, Morganti, Wong, Krause, Power, \& Hardcastle}]{Stewart_2025}
Stewart, G. S.~C., Shabala, S.~S., Yates-Jones, P.~M., {et~al.} in prep., Publications of the Royal Astronomical Society of Australia

\bibitem[{Sutherland \& Bicknell(2007)}]{Sutherland_2007}
Sutherland, R.~S., \& Bicknell, G.~V. 2007, The Astrophysical Journal Supplement Series, 173, 37–69

\bibitem[{Taub(1948)}]{Taub_1948}
Taub, A.~H. 1948, Physical Review, 74, 328–334

\bibitem[{Turner(2018)}]{Turner_2018}
Turner, R.~J. 2018, Monthly Notices of the Royal Astronomical Society, 476, 2522–2529

\bibitem[{Turner \& Shabala(2015)}]{Turner_2015}
Turner, R.~J., \& Shabala, S.~S. 2015, The Astrophysical Journal, 806, 59

\bibitem[{Turner \& Shabala(2023)}]{Turner_Shabala_2023}
---. 2023, Galaxies, 11, 87

\bibitem[{Turner {et~al.}(2023)Turner, Yates-Jones, Shabala, Quici, \& Stewart}]{Turner_etal_2023}
Turner, R.~J., Yates-Jones, P.~M., Shabala, S.~S., Quici, B., \& Stewart, G. S.~C. 2023, Monthly Notices of the Royal Astronomical Society, 518, 945

\bibitem[{Vernaleo \& Reynolds(2006)}]{Vernaleo_2006}
Vernaleo, J.~C., \& Reynolds, C.~S. 2006, The Astrophysical Journal, 645, 83–94

\bibitem[{Villaescusa-Navarro {et~al.}(2023)Villaescusa-Navarro, Genel, Anglés-Alcázar, Perez, Villanueva-Domingo, Wadekar, Shao, Mohammad, Hassan, Moser, Lau, Machado Poletti~Valle, Nicola, Thiele, Jo, Philcox, Oppenheimer, Tillman, Hahn, Kaushal, Pisani, Gebhardt, Delgado, Caliendo, Kreisch, Wong, Coulton, Eickenberg, Parimbelli, Ni, Steinwandel, La~Torre, Dave, Battaglia, Nagai, Spergel, Hernquist, Burkhart, Narayanan, Wandelt, Somerville, Bryan, Viel, Li, Irsic, Kraljic, Marinacci, \& Vogelsberger}]{Villaescusa_Navarro_2023}
Villaescusa-Navarro, F., Genel, S., Anglés-Alcázar, D., {et~al.} 2023, The Astrophysical Journal Supplement Series, 265, 54

\bibitem[{Virtanen {et~al.}(2020)Virtanen, Gommers, Oliphant, Haberland, Reddy, Cournapeau, Burovski, Peterson, Weckesser, Bright, {van der Walt}, Brett, Wilson, Millman, Mayorov, Nelson, Jones, Kern, Larson, Carey, Polat, Feng, Moore, {VanderPlas}, Laxalde, Perktold, Cimrman, Henriksen, Quintero, Harris, Archibald, Ribeiro, Pedregosa, {van Mulbregt}, \& {SciPy 1.0 Contributors}}]{Virtanen_Scipy}
Virtanen, P., Gommers, R., Oliphant, T.~E., {et~al.} 2020, Nature Methods, 17, 261

\bibitem[{Walg {et~al.}(2014)Walg, Achterberg, Markoff, Keppens, \& Porth}]{Walg_2014}
Walg, S., Achterberg, A., Markoff, S., Keppens, R., \& Porth, O. 2014, Monthly Notices of the Royal Astronomical Society, 439, 3969–3985

\bibitem[{Weinberger {et~al.}(2016)Weinberger, Springel, Hernquist, Pillepich, Marinacci, Pakmor, Nelson, Genel, Vogelsberger, Naiman, \& Torrey}]{Weinberger_2016}
Weinberger, R., Springel, V., Hernquist, L., {et~al.} 2016, Monthly Notices of the Royal Astronomical Society, 465, 3291–3308

\bibitem[{Yang \& Reynolds(2016)}]{Yang_Reynolds_2016}
Yang, K. H.-Y., \& Reynolds, C.~S. 2016, The Astrophysical Journal, 829, 90

\bibitem[{Yates {et~al.}(2018)Yates, Shabala, \& Krause}]{Yates_2018}
Yates, P.~M., Shabala, S.~S., \& Krause, M. G.~H. 2018, Monthly Notices of the Royal Astronomical Society, 480, 5286

\bibitem[{Yates-Jones {et~al.}(2021)Yates-Jones, Shabala, \& Krause}]{Yates_Jones_2021}
Yates-Jones, P.~M., Shabala, S.~S., \& Krause, M. G.~H. 2021, Monthly Notices of the Royal Astronomical Society, 508, 5239

\bibitem[{Yates-Jones {et~al.}(2023)Yates-Jones, Shabala, Krause, Hardcastle, {Mohd Noh Velast{\'i}n}, \& Stewart}]{YatesJones_2023}
Yates-Jones, P.~M., Shabala, S.~S., Krause, M. G.~H., {et~al.} 2023, Publications of the Astronomical Society of Australia, 40, e014

\bibitem[{Zanni {et~al.}(2005)Zanni, Murante, Bodo, Massaglia, Rossi, \& Ferrari}]{Zanni_2005}
Zanni, C., Murante, G., Bodo, G., {et~al.} 2005, Astronomy \&; Astrophysics, 429, 399–415

\bibitem[{Zovaro {et~al.}(2020)Zovaro, Sharp, Nesvadba, Kewley, Sutherland, Taylor, Groves, Wagner, Mukherjee, \& Bicknell}]{Zovaro_2020}
Zovaro, H. R.~M., Sharp, R., Nesvadba, N. P.~H., {et~al.} 2020, Monthly Notices of the Royal Astronomical Society, 499, 4940–4960

\end{thebibliography}

\appendix

\end{document}